\documentclass[american,nofootinbib]{revtex4-2}
\usepackage{lmodern}

\usepackage[T1]{fontenc}
\usepackage[latin9]{inputenc}
\usepackage{geometry}
\usepackage{color}
\usepackage{babel}
\usepackage{enumitem}
\usepackage{multirow}
\usepackage{amsmath}
\usepackage{amsthm}
\usepackage{amssymb}
\usepackage{graphicx}
\usepackage{dsfont}
\PassOptionsToPackage{normalem}{ulem}
\usepackage{ulem}
\usepackage[unicode=true,pdfusetitle,
 bookmarks=true,bookmarksnumbered=true,bookmarksopen=true,bookmarksopenlevel=1,
 breaklinks=false,pdfborder={0 0 0},pdfborderstyle={},backref=false,colorlinks=false]
 {hyperref}
\hypersetup{
 pdfborderstyle=}

\makeatletter

\providecommand{\tabularnewline}{\\}

\makeatother

\begin{document}
\title{Adiabatic radial perturbations of relativistic stars: analytic solutions
to an old problem}
\author{Paulo Luz}
\email{paulo.luz@tecnico.ulisboa.pt}

\affiliation{Centro de Astrof\'isica e Gravita\c{c}\~ao - CENTRA, Departamento de F\'isica,
Instituto Superior T\'ecnico - IST, Universidade de Lisboa - UL, Av. Rovisco
Pais 1, 1049-001 Lisboa, Portugal,}
\affiliation{Departamento de Matem\'atica, ISCTE - Instituto Universit\'ario de Lisboa,
Portugal}
\author{Sante Carloni}
\email{sante.carloni@unige.it}

\affiliation{Institute of Theoretical Physics, Faculty of Mathematics and Physics,
Charles University, Prague, V Hole\v{s}ovi\v{c}k\'ach 2, 180 00
Prague 8, Czech Republic,}
\affiliation{DIME, Universit\`a di Genova, Via all'Opera Pia 15, 16145 Genova,
Italy,}
\affiliation{INFN Sezione di Genova, Via Dodecaneso 33, 16146 Genova, Italy}

\begin{abstract}
We present a new system of equations that fully characterizes adiabatic,
radial perturbations of perfect fluid stars within the theory of general
relativity. The properties of the system are discussed, and, provided that the
equilibrium spacetime verifies some general regularity conditions,
analytical solutions for the perturbation variables are found. As
illustrative examples, the results are applied to study perturbations
of selected classical exact spacetimes, and the first oscillation eigenfrequencies
are computed. Exploiting the new formalism, we derive an upper bound for the maximum compactness of stable,
perfect fluid stars, which is equation-of-state-agnostic and significantly smaller than the
Buchdahl bound.
\end{abstract}
\maketitle

\section{Introduction}

Relativistic compact stellar objects are among the
most complex and, at the same time, most fascinating gravitational
systems. Similarly to black holes, these objects represent strong
gravity systems.  However, they are fundamentally different in that
the nature and behavior of matter play a prominent role in their structure
and evolution. This fact makes their
theoretical description particularly challenging. The
complexities associated with these objects explain, on the one hand,
the predominant use of numerical techniques and, on the other, the
necessity to develop and apply perturbative approaches
to understand their properties and dynamics. 

The first attempt at the description of perturbations of stars in
general relativity was made by Chandrasekhar in 1964~\citep{Chandrasekhar_1964_PRL,Chandrasekhar_1964_ApJ}.
In those works, Chandrasekhar focused on understanding the behavior of adiabatic,
radial perturbations and developed an integro-differential equation,
the so-called Chandrasekhar radial pulsation equation, to describe
this somewhat simpler type of perturbation. 

The Chandrasekhar equation has had a crucial influence on the subsequent studies on the dynamical behavior of
perturbations of self-gravitating, massive compact objects, and various methods were developed to compute the
oscillation eigenfrequencies directly from the pulsation equation~\citep{Bardeen_Thorne_Meltzer_1966}. Various
reformulations of Chandrasekhar's original equation have been proposed to ease the numerical treatment of
adiabatic, radial oscillations of self-gravitating fluids with realistic equations of
state (see e.g.~\citep{Chanmugam_1977, Vath_Chanmugam_1992, Gondek_Haensel_Zdunik_1997, Gondek_Zdunik_1999, Kokkotas_Ruoff_2001}).
However, to our knowledge, no work has so far tackled one of the main
limitations of Chandrasekhar's approach, i.e., the issue of gauge
dependence. Indeed, the predictions of the Chandrasekhar equation on the stability of a given matter configuration
and on the very behavior of its perturbations are intrinsically associated with the specific coordinate system considered
 and, therefore,  conditioned by the choice of the gauge.

Recently, the authors of this paper derived a completely covariant
and gauge-invariant theory of perturbations for static,
Locally Rotationally Symmetric of class II (LRS II) spacetimes~\cite{Luz_Carloni_2024a}.
LRS spacetimes are characterized by local rotational
symmetry about a given spatial direction, and their nonvortical subclass,
known as LRS II in the case of perfect fluid sources, can be proven
to contain all isotropic non-rotating spacetimes
suitable to describe stellar compact objects and even some slowly rotating
ones. The new perturbation framework is based on
the so-called 1+1+2 covariant formalism~\cite{Clarkson_Barrett_2003,Clarkson_2007}. This formalism is based on
the procedure of covariant spacetime decomposition, such that at, each
point, the spacetime is foliated by surfaces orthogonal to two vector
fields, a timelike, and a spacelike vector field. 
Using this covariant decomposition, it is possible to characterize different aspects of  the geometry of LRS II spacetimes and the thermodynamics of
the matter fields that permeate them in a geometrically clear and physically meaningful way~(see, e.g., Refs.~ \cite{Marklund_2003,Bradley_et_al_2012,Carloni_Vernieri_2018a, Carloni_Vernieri_2018b, Luz_Carloni_2019,
Tornkvist_Bradley_2019,Tornkvist_Bradley_2019,Naidu_Carloni_Dunsby_2021,Naidu_Carloni_Dunsby_2022,Rosa_Carloni_2023,Campbell_et_al_2024}).

The aim of this paper is to apply the covariant gauge-invariant perturbation
theory of Ref.~\cite{Luz_Carloni_2024a} to study
adiabatic, radial perturbations of static, compact stellar objects
composed of a perfect fluid.
In the body of the text, we will, however, write the
perturbation equations immediately in a non-covariant way from the
classical point of view of an observer locally comoving with the matter and
use the circumferential radius to identify points within the star. There are essentially two reasons for this 
choice. First, we aim to provide a set of equations that can be applied 
without any specific knowledge of the covariant formalisms. Second, the above setting 
is usually considered in the standard metric-based
description of this type of perturbation, providing a clear, familiar
interpretation of the quantities and the equations. We will then present
exact solutions for the perturbations of well-known models for equilibrium
spacetimes in a gauge-invariant way.

The paper is organized as follows. In Section \ref{sec:adiabatic-radial-perturbations},
we will present the equations describing adiabatic, radial perturbations,
and propose an algorithm to obtain power
series exact solutions of the system. We also give
a lower bound for the minimum eigenvalue of the system.
In Section \ref{sec:Perturbations_classical_solutions}, we apply
these methods to some classical solutions, namely, will consider the
interior Schwarzschild, Tolman IV, Kuchowicz 2-III and Heintzmann
IIa (as cataloged in~\citep{Delgaty_Lake_1998}) spacetimes, and
present the behavior of the first eigenfunctions that characterize
the perturbations in the frame of the comoving observer. In Section~\ref{sec:Maximum_compactness},
using results for the perturbation of the interior Schwarzschild solution, we conjecture
on a general upper bound for instability of static,
stellar compact objects composed of a perfect fluid,
which is independent of the information on the equation of state of the
perturbed matter fluid. We then draw conclusions in
Section~\ref{sec:Conclusions}. The paper also contains four appendices. In Appendix~\ref{sec:Appendix_1p1p2_decomposition},
we introduce the general definitions for the 1+1+2 covariant quantities.
In Appendix~\ref{sec:Appendix_Scalar_perturbation_equations}, we
give the general covariant, gauge-invariant perturbation equations
for adiabatic, radial perturbations found from the 1+1+2 formalism.
In Appendix~\ref{sec:Appendix_Pulsation_equation}, we show how the
original Chandrasekhar radial pulsation equation can be recovered
from the new gauge-invariant equations, demonstrating their equivalence
in the considered coordinate system, and in Appendix~\ref{sec:Appendix_Matrix_A}
we present the general intermediate matrix for the power series solutions
of the system.

Throughout the article, we will work in the geometrized unit system
where $8\pi G=c=1$, and consider the metric signature $(-+++)$.

\section{\label{sec:adiabatic-radial-perturbations}Adiabatic radial perturbations}

The Einstein field equations (EFE) are a set of nonlinear partial differential
equations that are manifestly difficult to solve for general source
matter fields. To circumvent these difficulties, it is helpful to devise
perturbation schemes to linearize the field equations and study the
behavior of small deviations from equilibrium solutions.
In Ref.~\cite{Luz_Carloni_2024a}, a general set of covariant
gauge-invariant equations was derived that can characterize linear perturbations
of static, spatially compact, spherically symmetric solutions of the
Einstein field equations. We will call such solutions ``stars''
because of their most immediate physical application. The gauge-invariant
equations for linear perturbations were written in the language of
the 1+1+2 covariant decomposition formalism, which relates the geometry
of the spacetime and the properties of the matter fluid with the kinematical
quantities that characterize two sets of congruences: a timelike congruence
and a spacelike congruence. As mentioned above, however,
we will relate all quantities directly with the metric tensor and
the matter fluid variables. Nonetheless, since the 1+1+2 variables have intrinsic
physical meaning, they are important to interpret the results. The reader can find the
basic definitions of the 1+1+2 formalism
in Appendix~\ref{sec:Appendix_1p1p2_decomposition}.

The evolution of general perturbations of stars is a compelling yet
remarkably complicated problem in gravitation theory. Here, we will
focus on studying adiabatic, radial perturbations of stars with a
perfect fluid source. The article is intended to be self-contained. Nonetheless, some general properties will be simply stated, and we redirect the reader to Ref.~\cite{Luz_Carloni_2024a} for technical details.

\subsection{The equilibrium spacetime and the perturbation variables\label{subsec:Background_spacetime}}

We will assume that the equilibrium background spacetime
is static and spherically symmetric, such that it can be
characterized by a line element of the form
\begin{equation}
ds_{0}^{2}=- \left(g_{0}\right)_{tt} dt^{2}+\left(g_{0}\right)_{rr} dr^{2}+r^{2}d\Omega^{2}\,,\label{eq:general_static_line_element}
\end{equation}
where $d\Omega^{2}$ represents the natural line element for the unit
2-sphere, $\left(t,r\right)$ are the standard Schwarzschild coordinates
measured by an observer at spatial infinity, and the metric
components $\left(g_{0}\right)_{tt}$ and $\left(g_{0}\right)_{tt}$
are functions of the circumferential radius, $r$, only. 
We will consider the setup where two solutions of the EFE are smoothly matched at a common timelike hypersurface, such that the interior of the star is described by a static, spatially compact solution with a perfect fluid source, with $r\in\left[0,r_{b}\right]$, where $r=r_{b}$ defines the boundary of the star, and the exterior spacetime, with $r>r_{b}$, is described by a radial branch of the vacuum Schwarzschild solution with no event horizons.
The metric coefficients $\left(g_{0}\right)_{tt}$ and $\left(g_{0}\right)_{rr}$ are determined by
the EFE with a zero cosmological constant
\begin{equation}
R_{\alpha\beta}-\frac{1}{2}g_{\alpha\beta}R
=T_{\alpha\beta}\,,
\end{equation}
where $R_{\alpha\beta}:=R_{\alpha\mu\beta}{}^{\mu}$ represents the
Ricci tensor, $R:=R_{\mu}{}^{\mu}$ the Ricci scalar, 
and $T_{\alpha\beta}$ the metric stress-energy
tensor described by a perfect fluid source, i.e. 
\begin{equation}
T_{\alpha\beta}=\left(\mu_{0}+p_{0}\right)(u_{0})_{\alpha}(u_{0})_{\beta}+p_{0}\left(g_{0}\right)_{\alpha\beta}\,,\label{T0Ord}
\end{equation}
where $u_0$ is the 4-velocity of the elements of volume of the fluid, and $\mu_{0}$ and $p_{0}$
are, respectively, the energy density and the isotropic pressure of
the matter fluid. Here and in the following, we will use the subscript ``0'' to explicitly
refer to quantities of the equilibrium spacetime.

It is useful to introduce the following scalar functions
\begin{equation}
\begin{aligned}\phi_{0} & =\frac{2}{r\sqrt{\left(g_{0}\right)_{tt} }}\,,\\
\mathcal{A}_{0} & =\frac{1}{2 \left(g_{0}\right)_{tt} \sqrt{\left(g_{0}\right)_{rr} }}\frac{d\left(g_{0}\right)_{tt}}{dr}\,,\\
\mathcal{E}_{0}= & \frac{1}{3}\mu_{0}+p_{0}-\mathcal{A}_{0}\phi_{0}\,.
\end{aligned}
\label{eq:Background_phi_A_E}
\end{equation}
The function $\phi_{0}$ represents the spatial expansion of the normalized
radial gradient vector field, $\mathcal{A}_{0}$
is the radial component of the 4-acceleration of the elements of volume
of the fluid, and $\mathcal{E}_{0}$ is the pure radial component
of the electric part of the Weyl tensor, Eq.~\eqref{Def_eq:Weyl_tensor_electric},
which partially describes radial tidal forces.

To characterize the perturbed spacetime with respect to the equilibrium background unambiguously, 
we have to choose variables that are identification gauge-invariant, that is, 
variables that are independent of the choice of diffeomorphism between the equilibrium and the perturbed spacetimes.
 Following
the Stewart-Walker Lemma~\citep{Stewart_Walker_1974}, variables that vanish identically in the background 
spacetime are gauge-invariant. 

Since the equilibrium spacetime is assumed static, the proper time derivatives of covariantly defined quantities vanish 
in the background and can be used to characterize the perturbed spacetime. Indeed, to describe the perturbations, we 
will consider the gauge-invariant variables
\begin{equation}
\begin{aligned}\mathsf{m} & :=\dot{\mu}\,, & \mathsf{p} & :=\dot{p}\,, & \mathsf{A} & :=\dot{\mathcal{A}}\,, & \mathsf{F} & :=\dot{\phi}\,, & \mathsf{E} & :=\dot{\mathcal{E}}\,,
\end{aligned}
\label{eq:GI_dot_derivatives_definition}
\end{equation}
where the ``dot'' represents the proper time derivative of an observer locally comoving with the fluid.
In addition to the variables in Eq.~\eqref{eq:GI_dot_derivatives_definition},
we will also consider  two other gauge-invariant variables: the expansion scalar $\theta$ associated with
the integral curves of the $u$ vector field, and the nontrivial radial
component of the shear tensor, $\Sigma$ (see Appendix \ref{sec:Appendix_1p1p2_decomposition} for details). The expansion 
scalar,
$\theta$, represents the fractional rate of change of the sectional
volume of the congruence associated with the vector field $u$ per
unit of proper time $\tau$, whereas $\Sigma$, together with $\theta$,
partially characterizes the proper time evolution of the radial inhomogeneity
of the matter fluid. Then, adiabatic, radial perturbations of perfect fluid
stars can be completely described in a frame comoving with the fluid,
by the set of gauge-invariant variables $\left\{ \mathsf{m},\mathsf{p},\mathsf{A},\mathsf{F},\mathsf{E},\theta,\Sigma\right\} $.

\subsection{Harmonic decomposition}

As is often the case in relativistic perturbation theory, the equilibrium solution is found
by considering highly symmetric setups. These symmetries can be
taken into account to transform the linearized system of partial differential
equations into a system of ordinary differential equations.
Indeed, in the case of a static, spherically symmetric equilibrium spacetime, for a suitable choice of frame and at
linear perturbation order, all quantities can be can be written in terms of the
Spherical Harmonics, $Y_{\ell m}$, and the eigenfunctions of the
Laplace operator in $\mathbb{R}$, $e^{i\upsilon\tau}$, such that,
a perturbed first-order scalar quantity $\chi$
can be written as 
\begin{equation}
\chi=\sum_{\upsilon}\left(\sum_{\ell=0}^{+\infty}\sum_{m=-\ell}^{\ell}\Psi_{\chi}^{\left(\upsilon,\ell\right)}Y_{\ell m}\right)e^{i\upsilon\tau}\,,
\end{equation}
where $\sum_{\upsilon}$ represents either a discrete sum or an integral
in $\upsilon$, depending on the boundary conditions of the problem,
$\tau$ is the proper time measured by an observer comoving with the
fluid in the background spacetime, $\upsilon$ represent
the eigenfrequencies, and $\Psi_{\chi}^{\left(\upsilon,\ell\right)}$
are the harmonic coefficients, which depend only on the radial coordinate
$r$.

In the case of isotropic perturbations, all coefficients $\Psi_{\chi}^{\left(\upsilon,\ell\right)}$
with $\ell\geq1$ are identically zero, that is,
dipole and higher-order angular multipoles must be trivial, otherwise
the perturbations would induce preferred directions in the system.
Moreover, it was rigorously shown in Ref.~\cite{Luz_Carloni_2024a}
that, under certain regularity conditions,
for this type of perturbation, the eigenfrequencies are such
that $\upsilon^{2}$ are countable, real, simple, have a minimum, and
are unbounded from above. Therefore, in the study of adiabatic, isotropic
perturbations a gauge-invariant, first-order scalar quantity $\chi$
can be decomposed as 
\begin{equation}
\chi=\sum_{\upsilon^{2}=\left\{ \upsilon_{0}^{2},\upsilon_{1}^{2},...\right\} }\Psi_{\chi}^{\left(\upsilon\right)}Y_{00}e^{i\upsilon\tau}\,,\label{eq:Radial_Adiabatic_general_harmonic_decompostion_upsilon}
\end{equation}
where the radial harmonic coefficients depend on the radial coordinate
only: $\Psi_{\chi}^{\left(\upsilon\right)}=\Psi_{\chi}^{\left(\upsilon\right)}\left(r\right)$,
and we have omitted the (trivial) dependency on $\ell$ to lighten the notation.

Alternatively, instead of considering the proper time $\tau$, we
can consider the time coordinate $t$. In that case, given the equation (see Ref.~\cite{Luz_Carloni_2024a})
\begin{equation}
\upsilon\left(r\right)=\frac{\lambda}{\sqrt{\left(g_{0}\right)_{tt} }}\,,
\label{eq:Radial_Adiabatic_eigenfrequencies_v-lambda_relation}
\end{equation}
relating the eigenfrequencies, $\upsilon$, measured by an observer
comoving with the fluid, with the constant eigenfrequencies $\lambda$,
measured by a static observer at spatial infinity, a gauge-invariant,
first-order scalar quantity $\chi$ can be equivalently given by Eq.~\eqref{eq:Radial_Adiabatic_general_harmonic_decompostion_upsilon}
or 
\begin{equation}
\chi=\sum_{\lambda^{2}=\left\{ \lambda_{0}^{2},\lambda_{1}^{2},...\right\} }\Psi_{\chi}^{\left(\lambda\right)}Y_{00}e^{i\lambda t}\,.
\end{equation}

\subsection{\label{Pert_Eq.}Gauge-invariant equation
of state and  perturbation equations}

The perturbation variables proposed previously in this section completely describe the dynamical evolution of the
adiabatically, radially perturbed spacetime. Nonetheless, to close the system, we need to characterize the perturbed
matter fluid by providing an equation of state. 
As a simplifying assumption, we will consider that the perturbed
matter fluid verifies a barotropic equation of state, such that 
\begin{equation}
p=f\left(\mu\right)\,,
\end{equation}
where $f$ is assumed to be twice differentiable in an open neighborhood
of $\mu_{0}$. Then, at linear order 
\begin{equation}
\mathsf{p}\approx f'\left(\mu_{0}\right)\mathsf{m}\,,
\label{eq:barotropic_eos_approx}
\end{equation}
where prime represents derivative with respect to the function's parameter,
so that $f'\left(\mu_{0}\right)$ represents the square of the adiabatic speed
of sound in the perturbed fluid, to be assumed nonvanishing in the
interior of the perturbed star. Notice that the function $f$ does
not have to be equal to the equation of state of the equilibrium configuration.
Indeed, the equilibrium fluid is not even required to verify a barotropic
equation of state. This freedom is physically relevant, as it was noted,
for instance, in Refs.~\citep{Gondek_Haensel_Zdunik_1997, Meltzer_Thorne_1966} where distinct adiabatic
indexes were considered for the equilibrium and the perturbed star.
Moreover, as we will see, in the case of an interior Schwarzschild background 
spacetime, the choice of $f$ is instrumental in analyzing its stability. 

In Appendix~\ref{sec:Appendix_Scalar_perturbation_equations} we
list the covariant, nontrivial perturbation equations for the variables
$\left\{ \mathsf{m},\mathsf{p},\mathsf{A},\mathsf{F},\mathsf{E},\theta,\Sigma\right\} $,
that characterize adiabatic, radial perturbations. Breaking
covariance, in the Schwarzschild coordinate system $\left(t,r\right)$
and considering the harmonic decomposition described in the previous
subsection, the radial coefficients associated with those variables
verify the following system of differential equations~\cite{Luz_Carloni_2024a}

\begin{equation}
\frac{d\Psi_{\mathsf{p}}^{\left(\upsilon\right)}}{dr}+\frac{4\mathcal{A}_{0}}{r\phi_{0}}\left(1+\frac{1}{3f'\left(\mu_{0}\right)}\right)\Psi_{\mathsf{p}}^{\left(\upsilon\right)}=-\frac{2\left(\mu_{0}+p_{0}\right)}{r\phi_{0}}\left(\Psi_{\mathsf{A}}^{\left(\upsilon\right)}+\mathcal{A}_{0}\Psi_{\Sigma}^{\left(\upsilon\right)}\right)\,,
\label{eq:Comoving_Radial_Adiabatic_pdot_hat}
\end{equation}
\begin{equation}
\frac{d\Psi_{\mathsf{A}}^{\left(\upsilon\right)}}{dr}+\left(\frac{6\mathcal{A}_{0}}{r\phi_{0}}-\frac{1}{r}\right)\Psi_{\mathsf{A}}^{\left(\upsilon\right)}=\frac{2\mathcal{E}_{0}}{r\phi_{0}\left(\mu_{0}+p_{0}\right)f'\left(\mu_{0}\right)}\Psi_{\mathsf{p}}^{\left(\upsilon\right)}-\frac{3}{r\phi_{0}}\left(\upsilon^{2}+\mathcal{A}_{0}^{2}+\frac{1}{3}\mu_{0}-2\mathcal{E}_{0}\right)\Psi_{\Sigma}^{\left(\upsilon\right)}\,,\label{eq:Comoving_Radial_Adiabatic_Adot_hat}
\end{equation}
\begin{equation}
\begin{aligned}\frac{d\Psi_{\Sigma}^{\left(\upsilon\right)}}{dr}+\left(\frac{3}{r}-\frac{4\mathcal{A}_{0}}{3r\phi_{0}f'\left(\mu_{0}\right)}\right)\Psi_{\Sigma}^{\left(\upsilon\right)} & =\frac{2}{3\left(\mu_{0}+p_{0}\right)f'\left(\mu_{0}\right)}\left[\left(\frac{f''\left(\mu_{0}\right)}{f'\left(\mu_{0}\right)}+\frac{1}{\mu_{0}+p_{0}}\right)\frac{d\mu_{0}}{dr}+\frac{1}{r\phi_{0}}\left(\frac{4}{3f'\left(\mu_{0}\right)}+2\right)\mathcal{A}_{0}\right]\Psi_{\mathsf{p}}^{\left(\upsilon\right)}\\
 & +\frac{4}{3r\phi_{0}f'\left(\mu_{0}\right)}\Psi_{\mathsf{A}}^{\left(\upsilon\right)}\,,
\end{aligned}
\label{eq:Comoving_Radial_Adiabatic_sigma_hat}
\end{equation}
and the constraints

\begin{equation}
\left(\upsilon^{2}+\mathcal{A}_{0}\phi_{0}+\mathcal{A}_{0}^{2}-p_{0}\right)\left(\frac{2}{3}\Psi_{\theta}^{\left(\upsilon\right)}-\Psi_{\Sigma}^{\left(\upsilon\right)}\right)=\Psi_{\mathsf{p}}^{\left(\upsilon\right)}-\phi_{0}\Psi_{\mathsf{A}}^{\left(\upsilon\right)}\,,
\label{eq:Comoving_Radial_Adiabatic_vsquare_constraint}
\end{equation}

\begin{equation}
\Psi_{\mathsf{E}}^{\left(\upsilon\right)}=\mathcal{E}_{0}\left(\frac{3}{2}\Psi_{\Sigma}^{\left(\upsilon\right)}+\frac{\Psi_{\mathsf{p}}^{\left(\upsilon\right)}}{f'\left(\mu_{0}\right)\left(\mu_{0}+p_{0}\right)}\right)-\frac{1}{2}\left(\mu_{0}+p_{0}\right)\Psi_{\Sigma}^{\left(\upsilon\right)}\,,
\end{equation}

\begin{equation}
\Psi_{\mathsf{F}}^{\left(\upsilon\right)}=\left(\frac{1}{2}\phi_{0}-\mathcal{A}_{0}\right)\left(\frac{2\Psi_{\mathsf{p}}^{\left(\upsilon\right)}}{3f'\left(\mu_{0}\right)\left(\mu_{0}+p_{0}\right)}+\Psi_{\Sigma}^{\left(\upsilon\right)}\right)\,,
\end{equation}
\begin{equation}
\Psi_{\mathsf{m}}^{\left(\upsilon\right)}=-\left(\mu_{0}+p_{0}\right)\Psi_{\theta}^{\left(\upsilon\right)}\,,\label{eq:Comoving_Radial_Adiabatic_mudot_constraint}
\end{equation}
\begin{equation}
\Psi_{\mathsf{p}}^{\left(\upsilon\right)}=f'\left(\mu_{0}\right)\Psi_{\mathsf{m}}^{\left(\upsilon\right)}\,,
\label{eq:Comoving_Radial_Adiabatic_mudot_pdot_eos}
\end{equation}
where Eq.~\eqref{eq:Comoving_Radial_Adiabatic_mudot_pdot_eos} follows from Eq.~\eqref{eq:barotropic_eos_approx}. The constraint equation equation~\eqref{eq:Comoving_Radial_Adiabatic_vsquare_constraint} is not propagated, therefore it cannot be used to reduce the size the system of differential equations~\eqref{eq:Comoving_Radial_Adiabatic_pdot_hat}--\eqref{eq:Comoving_Radial_Adiabatic_sigma_hat}. In fact, it is straightforward to show that Eqs.~\eqref{eq:Comoving_Radial_Adiabatic_pdot_hat}--\eqref{eq:Comoving_Radial_Adiabatic_vsquare_constraint} imply Eq.~\eqref{eq:Radial_Adiabatic_eigenfrequencies_v-lambda_relation}.

To select the physically acceptable solutions and formalize the boundary
value problem, we impose the following boundary conditions:
\begin{enumerate}[label=(\roman*)]
\item \label{enu:general_boundary_condition_1}the energy density and the
pressure perturbations at the center of the star, $r=0$, must be
finite in a neighborhood of the initial instant;
\item \label{enu:general_boundary_condition_2}the interior perturbed spacetime can
be smoothly matched to an exterior vacuum Schwarzschild spacetime
at a timelike hypersurface, the boundary of the star. 
\end{enumerate}
From the point of view of the comoving observer, the boundary condition~\ref{enu:general_boundary_condition_2}
implies that the pressure of the perturbed fluid is identically zero
at all times at the hypersurface and so, $\mathsf{p}$, hence $\Psi_{\mathsf{p}}^{\left(\upsilon\right)}$,
is also identically zero at the boundary, that is 
\begin{equation}
\Psi_{\mathsf{p}}^{\left(\upsilon\right)}\left(r_{b}\right)=0\,.
\end{equation}

Before proceeding, we remark that, in general, the coordinate system
in the background spacetime and that of the perturbed spacetime are
not necessarily the same: any smooth mapping can be considered. Since
the perturbation variables are gauge-invariant, that choice does not
affect the results. In the particular case of isotropic perturbations,
the Schwarzschild coordinate system can always be defined since there
is no gravitational wave emission, and the spacetime is asymptotically
flat, hence the time coordinate $t$ and circumferential radius $r$
are equally defined by an observer at spatial infinity in both spacetimes.
In particular, the adoption of this coordinate system is useful to
compare our approach with the classical results of Chandrasekhar,
found from metric based perturbation theory~\citep{Chandrasekhar_1964_PRL,Chandrasekhar_1964_ApJ}.
In Appendix~\ref{sec:Appendix_Pulsation_equation} we show explicitly
that Chandrasekhar's second-order radial pulsation equation follows
from the system above by relating the kinematical quantities with
the radial displacement parameter and its derivatives.

\subsection{Analytic solutions}

The system~\eqref{eq:Comoving_Radial_Adiabatic_pdot_hat}--\eqref{eq:Comoving_Radial_Adiabatic_mudot_pdot_eos}
with boundary conditions~\ref{enu:general_boundary_condition_1}
and \ref{enu:general_boundary_condition_2} completely characterizes
adiabatic, radial perturbations of a star composed of a perfect fluid.
Specifying the background spacetime and the equation of state of the
perturbed fluid, numerical methods can be used to find approximate
solutions. Nonetheless, contrary to the original form of the second-order
Chandrasekhar's pulsation equation~\citep{Chandrasekhar_1964_PRL,Chandrasekhar_1964_ApJ},
or the associated first-order realizations of Refs.~\citep{Chanmugam_1977, Gondek_Haensel_Zdunik_1997},
it is possible, under rather general conditions, to find analytic
solutions for the perturbations using standard theory
of systems of linear ordinary differential equations.

To find analytic solutions for the system~\eqref{eq:Comoving_Radial_Adiabatic_pdot_hat}--\eqref{eq:Comoving_Radial_Adiabatic_mudot_pdot_eos},
we will further impose that the following regularity constraints hold:
\begin{itemize}
\item \label{enu:regularity_conditions}the equilibrium fluid verifies the
weak energy condition; 
\item the background spacetime is a solution of the Tolman-Oppenheimer-Volkoff
equation for real analytic, nontrivial energy density and pressure
functions for the whole range within the equilibrium star; 
\item the square of the speed of sound of the perturbed fluid, $f'$, is
positive and real analytic in the interior and at the boundary of
the star.
\end{itemize}
Requiring real analytical background solutions is a rather strong
constraint. However, to our knowledge, all classical exact solutions
for compact astrophysical objects verify this hypothesis in some open 
neighborhood of the center of the star  ($r=0$). Hence, the following
results are appropriate for treating perturbations of physically relevant
setups. The radius of convergence of the power series, of course, may
or may not be greater than the radius of the equilibrium star. In the latter case, further treatment has to be carried out to extend the
solution to the boundary. Nonetheless, to prove the following results, it
suffices that the radius of convergence of the power series around
the center is nonzero.

Imposing the above conditions, the background EFE equations imply 
\begin{equation}
\begin{aligned}\phi_{0} & =\frac{2}{r}\sqrt{1-\frac{2M\left(r\right)}{r}}\,,\\
\mathcal{E}_{0} & =\frac{1}{3}\mu_{0}-\frac{2M\left(r\right)}{r^{3}}\,,\\
\mathcal{A}_{0}\phi_{0} & =p_{0}+\frac{2M\left(r\right)}{r^{3}}\,,
\end{aligned}
\end{equation}
where 
\begin{equation}
M\left(r\right):=\frac{1}{2}\int_{0}^{r}\mu_{0}x^{2}dx\,,
\end{equation}
is usually called the mass function. If the functions $\mu_{0}$ and
$p_{0}$ verify the weak energy condition and are real analytic within
the star, so are the functions $\mathcal{A}_{0}$ and $\mathcal{E}_{0}$,
and $\phi_{0}$ has a simple pole at the center, $r=0$, but is otherwise
real analytic in the interior and boundary of the star. Therefore,
in the considered setup, Eqs.~\eqref{eq:Comoving_Radial_Adiabatic_pdot_hat}--\eqref{eq:Comoving_Radial_Adiabatic_sigma_hat}
form a system of ordinary differential equations with real analytic
coefficients around $r=0$, with a simple pole at $r=0$, and solutions
can be found around the singular point.

Consider the system~\eqref{eq:Comoving_Radial_Adiabatic_pdot_hat}--\eqref{eq:Comoving_Radial_Adiabatic_sigma_hat}
in matrix form:
\begin{equation}
\frac{d\mathds{W}}{dr}=\left(r^{-1}\mathds{R}+\Theta\right)\mathds{W}\,,\label{eq:Radial_Adiabatic_Comoving_matrix_form}
\end{equation}
where
\begin{equation}
\begin{aligned}\mathds{W} & =\left[\begin{array}{c}
\Psi_{\mathsf{p}}^{\left(\lambda\right)}\\
\Psi_{\mathsf{A}}^{\left(\lambda\right)}\\
\Psi_{\Sigma}^{\left(\lambda\right)}
\end{array}\right]\,, &  &  & \mathds{R} & =\left[\begin{array}{lcr}
0 & 0 & 0\\
0 & 1 & 0\\
0 & 0 & -3
\end{array}\right]\,,\end{aligned}
\end{equation}
\begin{equation}
\begin{aligned}\Theta & =\frac{2}{r\phi_{0}}\left[\begin{array}{lcr}
-2\mathcal{A}_{0}\left(1+\frac{1}{3f'\left(\mu_{0}\right)}\right) & \quad-\left(\mu_{0}+p_{0}\right) & \quad-\left(\mu_{0}+p_{0}\right)\mathcal{A}_{0}\\
\\
\frac{\mathcal{E}_{0}}{f'\left(\mu_{0}\right)\left(\mu_{0}+p_{0}\right)} & \quad-3\mathcal{A}_{0} & \quad-\frac{3}{2}\left(\upsilon^{2}+\mathcal{A}_{0}^{2}+\frac{1}{3}\mu_{0}-2\mathcal{E}_{0}\right)\\
\\
\frac{3f''\left(\mu_{0}\right)r\phi_{0}\partial_{r}\mu_{0}+4\mathcal{A}_{0}}{9\left(\mu_{0}+p_{0}\right)\left[f'\left(\mu_{0}\right)\right]^{2}}+\frac{r\phi_{0}\partial_{r}\mu_{0}+2\mathcal{A}_{0}\left(\mu_{0}+p_{0}\right)}{3\left(\mu_{0}+p_{0}\right)^{2}f'\left(\mu_{0}\right)} & \quad\frac{2}{3f'\left(\mu_{0}\right)} & \quad\frac{2\mathcal{A}_{0}}{3f'\left(\mu_{0}\right)}
\end{array}\right]\,.\end{aligned}
\label{eq:Radial_Adiabatic_comoving_Theta_matrix}
\end{equation}
The regularity conditions that we have imposed on the thermodynamic
variables of the equilibrium configuration, $\mu_{0}$ and $p_{0}$,
and on the equation of state of the perturbed fluid, $f$, guarantee
that the matrix $\Theta$ is real analytic at $r=0$, in particular,
the fact that $\mu_{0}$ and $p_{0}$ are differentiable functions
guarantees that $r\phi_{0}$ does not vanish in the interior of the
star. From Eq.~\eqref{eq:Radial_Adiabatic_Comoving_matrix_form}
we conclude that $r=0$ is a regular singular point of the system.
However, of course, this does not imply that all solutions
must be singular at the center. To select only the physically acceptable
solutions, we will impose the boundary conditions~\ref{enu:general_boundary_condition_1}
and \ref{enu:general_boundary_condition_2}.

To solve the system of ordinary differential equations, we will follow
the formalism in Ref.~\citep{Coddington_Levinson_Book} and find
solutions in a power series form. Since the $\Theta$ matrix is assumed
to be real analytic at the center of the star, it can be expanded in a convergent
power series of the form 
\begin{equation}
\Theta\left(r\right)=\sum_{n=0}^{+\infty}\Theta_{n}r^{n}\,.
\label{eq:Theta_matrix_power_series}
\end{equation}
Then, the solution matrix $\mathds{W}$ can be written in a power series, guaranteed
to converge to the solution in a neighborhood of $r=0$. The radius
of convergence of the power series solution is equal, except for possibly
at $r=0$, to the radius of convergence of the power series of $\Theta$,
which, of course, depends on the considered background spacetime.

Before presenting the formal solutions, we remark that for a
general static, spherically symmetric spacetime, the general form
for the perturbative solutions can be rather complicated, however,
considering the regularity conditions imposed on $\mu_{0}$ and $p_{0}$,
and assuming $f'\left(\mu_{0}\right)$ is not zero in a neighborhood
of the center of the star, using the Tolman-Oppenheimer-Volkoff equations
it can be shown that various entries of the coefficient-matrices $\left\{ \Theta_{0},\Theta_{1},\Theta_{2},\Theta_{3}\right\}$  are zero, which greatly simplifies the general family of
solutions of the system. Then,  taking into consideration the regularity of the background solution, applying the method in Ref.~\citep{Coddington_Levinson_Book} yields 
\begin{equation}
\begin{aligned}\left[\begin{array}{c}
\Psi_{\mathsf{p}}^{\left(\lambda\right)}\\
\Psi_{\mathsf{A}}^{\left(\lambda\right)}\\
\Psi_{\Sigma}^{\left(\lambda\right)}
\end{array}\right] & =\left[\begin{array}{rcr}
-1 & \frac{12}{r}\left[\left(\Theta_{0}\right)_{12}\left(\Theta_{0}\right)_{23}-3\left(\Theta_{1}\right)_{13}\right] & \:0\\
0 & \:12\left(\Theta_{0}\right)_{23}\left[\left(\Theta_{1}\right)_{22}-\left(\Theta_{1}\right)_{33}\right]-18\left(\Theta_{2}\right)_{23}+4\left[\left(\Theta_{0}\right)_{23}\right]^{2}\left(\Theta_{0}\right)_{32}-\frac{12}{r^{2}}\left(\Theta_{0}\right)_{23} & r\\
0 & \frac{36}{r^{3}} & 0
\end{array}\right]\mathds{P}_{\mathds{W}}\left[\begin{array}{c}
c_{1}\\
c_{2}\\
c_{3}
\end{array}\right]\,,\end{aligned}
\label{eq:Radial_Adiabatic_comoving_matrix_system}
\end{equation}
where $c_{1}$, $c_{2}$ and $c_{3}$ are integration constants, which
might depend on the constant eigenfrequencies $\lambda$, defined
in Eq.~\eqref{eq:Radial_Adiabatic_eigenfrequencies_v-lambda_relation}.
The notation $\left(\Theta_{n}\right)_{ij}$ is to be interpreted
as the $ij$-entry of the $n$th order matrix coefficient of the power
series expansion of $\Theta$, Eq.~\eqref{eq:Theta_matrix_power_series}, and $\mathds{P}_{\mathsf{W}}$ is a real analytic
matrix, such that
\begin{equation}
\begin{aligned}\mathds{P}_{\mathds{W}}\left(r\right) & =\sum_{n=0}^{+\infty}\mathds{P}_{n}r^{n}\,,\\
\mathds{P}_{0} & =\mathds{I}_{3}\,,\\
\mathds{P}_{k} & =\frac{1}{k}\sum_{j=0}^{k-1}\mathds{A}_{k-1-j}\mathds{P}_{j}\,,\quad\text{for }k\geq1\,,
\end{aligned}
\label{eq:Radial_Adiabatic_comoving_recurrence_relation}
\end{equation}
where $\mathds{I}_{3}$ represents the $3\times3$ identity matrix, the matrix
$\mathds{A}$ is real analytic with the same radius of convergence as $\Theta$,
and $\mathds{A}_{n}$ represents the $n$th order matrix-coefficient of its power
series expansion, that is, $\mathds{A}\left(r\right)=\sum_{n=0}^{+\infty}\mathds{A}_{n}r^{n}$. Due to its size, the matrix $\mathds{A}$ is presented in Appendix~\ref{sec:Appendix_Matrix_A}.

Using $\mathds{P}_{0}=\mathds{I}_{3}$ in Eq.~\eqref{eq:Radial_Adiabatic_comoving_matrix_system},
we can directly compute the lower order coefficients of the power
series expansion of $\mathds{W}$. Imposing the boundary condition
at the center sets the coefficient $c_{2}$ to be zero, otherwise
the pressure would diverge at $r=0$ at all times. Then, we find 
\begin{equation}\label{Series_solution}
\left[\begin{array}{c}
\Psi_{\mathsf{p}}^{\left(\lambda\right)}\\
\Psi_{\mathsf{A}}^{\left(\lambda\right)}\\
\Psi_{\Sigma}^{\left(\lambda\right)}
\end{array}\right]=\left[\begin{array}{c}
-c_{1}+\mathcal{O}\left(r^{2}\right)\\
c_{3}r+\mathcal{O}\left(r^{3}\right)\\
\mathcal{O}\left(r^{2}\right)
\end{array}\right]\,.
\end{equation}
We see that the coefficient $c_{1}$ directly characterizes the behavior
of $\Psi_{\mathsf{p}}^{\left(\lambda\right)}$ at $r=0$ and that
both $\Psi_{\mathsf{A}}^{\left(\lambda\right)}$ and $\Psi_{\Sigma}^{\left(\lambda\right)}$
must vanish at the center. On the other hand, the coefficients $c_{1}$
and $c_{3}$ are not independent: considering the regularity of the
background spacetime and imposing the constraints~\eqref{eq:Comoving_Radial_Adiabatic_vsquare_constraint},
\eqref{eq:Comoving_Radial_Adiabatic_mudot_constraint} and \eqref{eq:Comoving_Radial_Adiabatic_mudot_pdot_eos}
leads to 
\begin{equation}
c_{3}\stackrel{r=0}{=}-\frac{c_{1}}{3\left(\mu_{0}+p_{0}\right)f'\left(\mu_{0}\right)}\left[ \frac{\lambda^{2}}{\left(g_{0}\right)_{tt}} +\frac{1}{3}\mu_{0}+\frac{3}{2}\left(\mu_{0}+p_{0}\right)f'\left(\mu_{0}\right)\right]\,,
\end{equation}
where all quantities on the right-hand side are to be evaluated at
$r=0$. Therefore, for each value of $\lambda^{2}$ there is a single
arbitrary parameter, either $c_{1}$ or $c_{3}$, to be characterized
by the initial perturbation at the center of the star.
That is, the
coefficients of the Fourier transform of the initial perturbation
provide the values of the independent parameters, setting which eigenmodes
are excited and the respective initial magnitude. In
what follows we will consider $c_{1}$ as the independent parameter.
Specifying the background spacetime, the equation of state of the
perturbed fluid and the values of the eigenfrequencies $\lambda$,
these results allow us to find regular analytic solutions for the
perturbations that verify the boundary conditions.

In addition to the analytical results for the eigenfunctions, in Ref.~\cite{Luz_Carloni_2024a}
it was possible to establish lower bounds for the square of the eigenfrequencies,
$\lambda^{2}$. In particular, if the regularity conditions in the
beginning of this subsection and the boundary conditions~\ref{enu:general_boundary_condition_1}
and \ref{enu:general_boundary_condition_2} hold, nontrivial $\mathcal{C}^{1}$
solutions of the system \eqref{eq:Comoving_Radial_Adiabatic_pdot_hat}--\eqref{eq:Comoving_Radial_Adiabatic_mudot_pdot_eos}
exist only if
\begin{equation}
\lambda^{2}\max_{r\in\left]0,r_{b}\right[} \left(g_{0}\right)_{tt} >-\max_{r\in\left]0,r_{b}\right[}\left[\frac{\mu_{0}+p_{0}}{\phi_{0}}\left(\frac{1}{2}\phi_{0}+2\mathcal{A}_{0}\right)+\frac{\mathcal{A}_{0}^{2}}{f'\left(\mu_{0}\right)}+\frac{r\phi_{0}\mathcal{A}_{0}}{2\left(\mu_{0}+p_{0}\right)}\frac{d\mu_{0}}{dr}\right]\,.\label{eq:Radial_Adiabatic_comoving_bound_freq}
\end{equation}
This result can be useful to determine numerically
the eigenfrequencies of the system, offering a baseline to search
for their values.

\section{\label{sec:Perturbations_classical_solutions}Perturbations and fundamental
eigenfrequencies of classic exact solutions}

As illustrative examples of the general results in the previous section,
we will study the properties of adiabatic, radial perturbations of
some classical solutions of perfect fluid stars within general relativity.
Since the independent parameter $c_{1}$ simply characterizes the
magnitude of a specific eigenfunction at $r=0$, without loss of generality,
in this section we will set $c_{1}=-1$.

In Table~\ref{table:Metric-coefficients},
considering a line element of the form~\eqref{eq:general_static_line_element},
we present the nontrivial metric coefficients of some classical, physically relevant solutions
of the Einstein field equations. In Table~\ref{table:Fundamental-modes}
we present the absolute value of the first three eigenfrequencies for specific models for the equilibrium background
spacetime. For the interior Schwarzschild solution, at linear order, the perturbed fluid	is completely characterized by a constant speed of sound, which must be provided as an extra parameter.
For the other models, we assume
that the equation of state of the perturbed fluid is the same as that of
the equilibrium setup. For all considered models, the eigenfrequencies take real values, therefore all equilibrium spacetimes represent stable configurations under adiabatic, radial perturbations.

\begin{figure}
	\centering\includegraphics[width=1\textwidth]{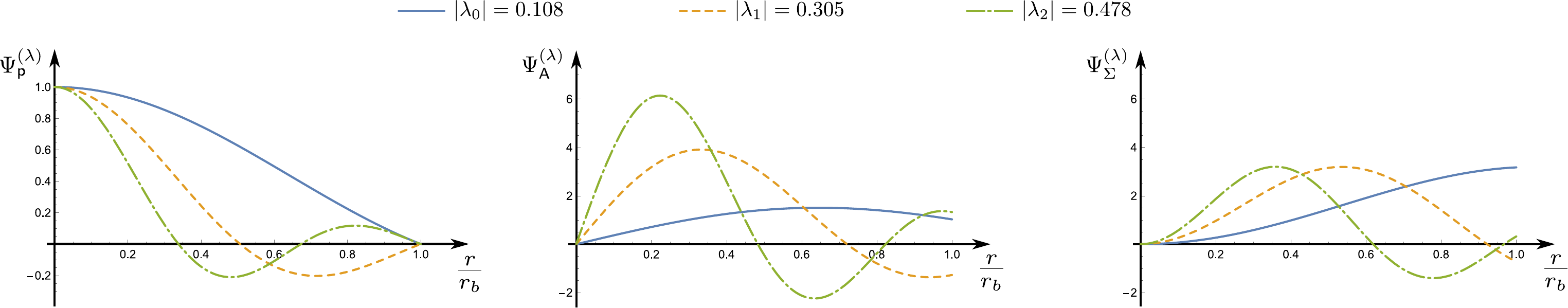}
	
	\caption{\label{fig:Eigenfunctions_Schwarzschild}Radial profile of the Fourier
	coefficients of the functions $\mathsf{p}$, $\mathsf{A}$, and $\Sigma$,
	associated with the eigenfrequencies presented in Table~\ref{table:Fundamental-modes}
	for the interior Schwarzschild spacetime. For all oscillation modes,
	it was assumed $c_{1}=-1$.}
\end{figure}

\begin{figure}
	\centering\includegraphics[width=1\textwidth]{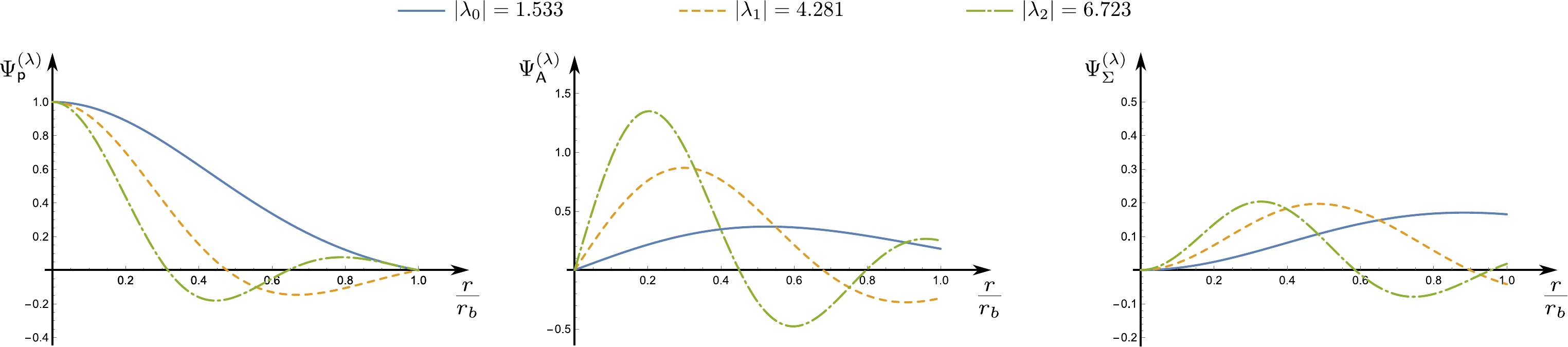}
	
	\caption{\label{fig:Eigenfunctions_TolmanIV}Radial profile of the Fourier
	coefficients of the functions $\mathsf{p}$, $\mathsf{A}$, and $\Sigma$,
	associated with the eigenfrequencies presented in Table~\ref{table:Fundamental-modes}
	for the Tolman IV spacetime. For all oscillation modes, it was assumed
	$c_{1}=-1$.}
\end{figure}

\begin{figure}
	\centering\includegraphics[width=1\textwidth]{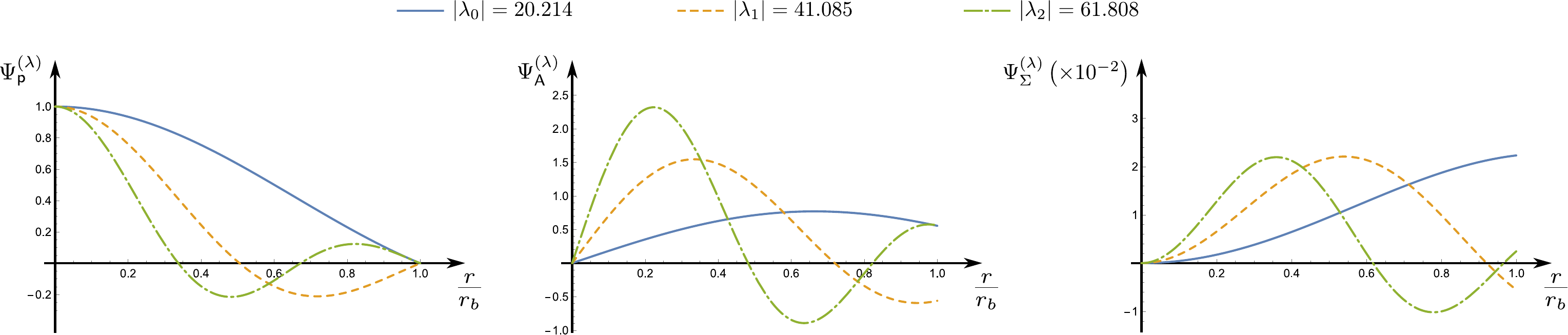}
	
	\caption{\label{fig:Eigenfunctions_Kuch2III}Radial profile of the Fourier
	coefficients of the functions $\mathsf{p}$, $\mathsf{A}$, and $\Sigma$,
	associated with the eigenfrequencies presented in Table~\ref{table:Fundamental-modes}
	for the Kuchowicz 2-III spacetime. For all oscillation modes, it was
	assumed $c_{1}=-1$.}
\end{figure}

\begin{figure}
	\centering\includegraphics[width=1\textwidth]{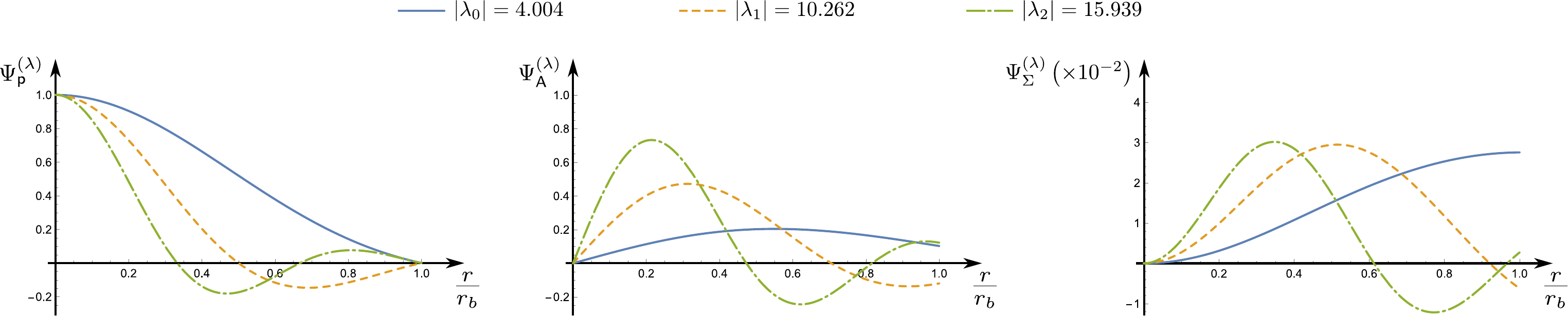}
	
	\caption{\label{fig:Eigenfunctions_HeintIIa}Radial profile of the Fourier
	coefficients of the functions $\mathsf{p}$, $\mathsf{A}$, and $\Sigma$,
	associated with the eigenfrequencies presented in Table~\ref{table:Fundamental-modes}
	for the Heintzmann IIa spacetime. For all oscillation modes, it was
	assumed $c_{1}=-1$.}
\end{figure}

In Figures~\ref{fig:Eigenfunctions_Schwarzschild}--\ref{fig:Eigenfunctions_HeintIIa}
we present the radial profile of the Fourier coefficients of the functions
$\mathsf{p}$, $\mathsf{A}$, and $\Sigma$, associated with the eigenfrequencies
presented in Table~\ref{table:Fundamental-modes} for the various
background spacetimes. Figures~\ref{fig:Eigenfunctions_Schwarzschild}--\ref{fig:Eigenfunctions_HeintIIa} highlight the expected behavior for the eigenfunctions. For a real-analytic background spacetime, in Ref.~\cite{Luz_Carloni_2024a} it was shown that the perturbations equations can be cast in the form of a Sturm-Liouville eigenvalue problem. Therefore,  in particular, the number of roots of the eigenfunctions is associated with the order of the associated eigenvalue in the sequence $\left( \lambda^2_n\right)_{n\in \mathbb{N}}$~.

\begin{table}[h]
\centering%
\begin{tabular}{|c|c|}
\hline 
Spacetime  & Non-trivial metric components\tabularnewline
\hline 
\hline 
\multirow{1}{*}{\,\,Interior Schwarzschild\,\,} & $\begin{aligned}\\
\left(g_{0}\right)_{tt} = & \left(3\sqrt{1-\frac{2M}{r_{b}}}-\sqrt{1-\frac{2Mr^{2}}{r_{b}^{3}}}\right)^{2}\\
\left(g_{0}\right)_{rr} = & \left(1-\frac{2Mr^{2}}{r_{b}^{3}}\right)^{-1}\\
\\
\end{aligned}
$\tabularnewline
\hline 
\multirow{1}{*}{Tolman IV} & $\begin{aligned}\\
\left(g_{0}\right)_{tt} = & B^{2}\left(\frac{r^{2}}{A^{2}}+1\right)\\
\left(g_{0}\right)_{rr} = & \frac{\frac{2r^{2}}{A^{2}}+1}{\left(1+\frac{r^{2}}{A^{2}}\right)\left(1-\frac{r^{2}}{R^{2}}\right)}\\
\\
\end{aligned}
$\tabularnewline
\hline 
\multirow{1}{*}{Kuch2 III} & $\begin{aligned}\\
\,\,\left(g_{0}\right)_{tt} = & Be^{\frac{Ar^{2}}{2}}\\
\,\,\left(g_{0}\right)_{rr} = & \left(r^{2}e^{-\frac{1}{2}Ar^{2}}\left[C-\frac{A}{2e}\text{Ei}\left(\frac{Ar^{2}}{2}+1\right)\right]+1\right)^{-1}\\
\\
\end{aligned}
$\tabularnewline
\hline 
\multirow{1}{*}{Heint IIa} & $\begin{aligned}\\
\left(g_{0}\right)_{tt} = & A^{2}\left(ar^{2}+1\right)^{3}\\
\left(g_{0}\right)_{rr} = & \left(1-\frac{3ar^{2}\left[c\left(4ar^{2}+1\right)^{-\frac{1}{2}}+1\right]}{2\left(ar^{2}+1\right)}\right)^{-1}\\
\\
\end{aligned}
$\tabularnewline
\hline 
\end{tabular}

\caption{\label{table:Metric-coefficients}Metric coefficients of classical
solutions of the Einstein field equations. The spacetimes are assumed
to be characterized by a line element of the form of Eq.~\eqref{eq:general_static_line_element}. We follow the naming conventions for the solutions
of Ref.~\citep{Delgaty_Lake_1998}.}
\end{table}

\begin{table}[h]
	\centering%
	\begin{tabular}{|c|c|c|c|c|}
	\hline 
	Spacetime  & Parameters  & $\left| \lambda_0\right|$  & $\left| \lambda_1\right|$  & $\left| \lambda_2\right|$\tabularnewline
	\hline 
	\hline 
	\multirow{1}{*}{\,\,Interior Schwarzschild\,\,} & $\,\,\left(M,r_{b},c_{s}^{2}\right) =\left( 0.1,1,0.1\right) \,\,$  & 0.108  & 0.305  &  0.478\tabularnewline
	\hline 
	Tolman IV  & $\left( A,B,R\right) =\left( 1,1,1.5\right) $  & 1.533  & 4.281  & 6.723\tabularnewline
	\hline 
	Kuch2 III  & $\left( A,B,C\right) =\left( 5,1,-3\right) $  &\,\, 20.214 \,\, &\,\, 41.085 \,\, & \,\,61.808\,\,\tabularnewline
	\hline 
	Heint IIa  & $\left( a,A,C\right) =\left( 1,1,1.5\right) $  &  4.004  & 10.262  & 15.939\tabularnewline
	\hline 
	\end{tabular}
	
	\caption{\label{table:Fundamental-modes}First absolute values of the eigenfrequencies,
	$\lambda$, rounded to three decimal places, for the equilibrium solutions
	in Table~\ref{table:Metric-coefficients} for specific values of
	the spacetime parameters. In all examples, the eigenfrequencies are real, hence all spacetimes are stable under adiabatic, radial perturbations.
	We follow the naming conventions for the solutions
	of Ref.~\citep{Delgaty_Lake_1998}.}
\end{table}

Except for the interior Schwarzschild solution, the results in Table~\ref{table:Fundamental-modes}
were compared with the predictions of the systems in Refs.~\citep{Chanmugam_1977, Gondek_Haensel_Zdunik_1997}.
Implementing a shooting method to solve numerically each of those 
systems for each equilibrium spacetime, all values for the fundamental
eigenfrequencies matched to the considered numerical accuracy. Moreover,
the general solutions found in Sec.~\ref{sec:adiabatic-radial-perturbations}
are regular around the center, and for the spacetime parameters in
Table~\ref{table:Fundamental-modes}, the solutions are exact and the
radius of convergence of the power series is greater than the radius
of the star. Hence, we can explicitly evaluate the boundary conditions
at any point within the star, in particular at the center and the
surface.

\section{\label{sec:Maximum_compactness}Bound on maximum compactness}

The dynamical stability of an equilibrium static, self-gravitating
fluid is a crucial problem in astrophysics. Finding that a given solution
of the Einstein field equations is dynamically unstable for some type
of perturbation implies that such a solution might not be suitable
to describe massive compact objects that are expected to be empirically
observable in the current universe. In that regard, considerable effort
has been devoted to finding an upper bound for the maximum compactness
of a star: the ratio between the gravitational mass and the circumferential 
radius of a star, beyond which it is unstable 
(cf., e.g., Refs.~\citep{Alho_et_al_2022, Alho_et_al_2024} for 
a review for perfect and non-perfect fluids, and Ref.~\cite{Masa_et_al_2023}
and references therein for the case of electrically charged self-gravitating fluids).
Nonetheless, the results in the literature are fundamentally 
connected to the equation of state of the equilibrium fluid and, to our knowledge, no universal
upper bound has been found for the maximum compactness of these types of objects.
In the particular case of perfect fluid stars, we can use the previous
results and the interior Schwarzschild solution to conjecture an equation
of state agnostic upper bound for the dynamical stability of this
type of objects.

The interior Schwarzschild spacetime represents a solution of general
relativity describing the interior of a star composed of a perfect
fluid with a constant energy density that can be smoothly matched with
an exterior vacuum Schwarzschild spacetime. This solution can be characterized
by a line element of the form~\eqref{eq:general_static_line_element},
with
\begin{equation}
\begin{aligned}\left(g_{0}\right)_{tt} = & \left(3\sqrt{1-\frac{2M}{r_{b}}}-\sqrt{1-\frac{2Mr^{2}}{r_{b}^{3}}}\right)^{2}\,,\\
\left(g_{0}\right)_{rr} = & \left(1-\frac{2Mr^{2}}{r_{b}^{3}}\right)^{-1}\,,
\end{aligned}
\end{equation}
where $M$ represents the ADM mass and $r_{b}$ the value of the circumferential
radius at the boundary of the star. Although the interior Schwarzschild
solution does not represent a physically reasonable configuration, this
solution is important from a conceptual point of view since it saturates
the Buchdahl bound for the maximum compactness:
\begin{equation}
\frac{M}{r_{b}}\leq\frac{4}{9}=0.\overline{4}\,,
\label{eq:Buchdahl_bound}
\end{equation}
hence, it can be thought of as the extreme scenario for a static, self-gravitating
perfect fluid. On this basis, this solution can be used to conjecture
the maximum compactness of a star composed of a perfect fluid, beyond
which it becomes dynamically unstable.

Considering the results of the previous section, we can study the
dynamical stability of the interior Schwarzschild solution by computing
the fundamental eigenfrequency for various values of the compactness
parameter $\frac{M}{r_{b}}$. Notice that, for an
interior Schwarzschild background spacetime with a specific value
of the compactness parameter, the value of $r_{b}$ does not affect
the values of the eigenfrequencies,  but only the radial profile of the
perturbation variables. Since the energy density of the background
is constant, at a linear level, the square of the speed of sound, $f'\left(\mu_{0}\right)$,
is a constant. Then, to infer the maximum compactness of a physical
star described by the interior Schwarzschild solution, we can impose
the causality condition, and consider the extreme scenario where  $f'\left(\mu_{0}\right)=c_{s}^{2}=1$, that
is, the square of the speed of sound of the perturbed star is the
vacuum speed of light. Applying the previous results, we find that
\begin{equation}
M/r_{b}=0.367\Rightarrow\lambda_{0}^{2}>0\,, \qquad M/r_{b}=0.368\Rightarrow\lambda_{0}^{2}<0\,.
\end{equation}
Following this reasoning, we extrapolate that a static, spherically
symmetric solution of the Einstein field equations with a perfect
fluid source is dynamically unstable if
\begin{equation}
\frac{M}{r_{b}}\gtrsim0.368\,.\label{eq:min_bound_unstable}
\end{equation}
The accuracy of the above estimate, of course, can be increased,
but this level of accuracy is unlikely to be achieved experimentally.
The result in Eq.~(\ref{eq:min_bound_unstable}) is
significantly smaller than the Buchdahl bound, Eq.~\eqref{eq:Buchdahl_bound}.
Moreover, it is slightly higher, but in line with the estimates
of Refs.~\citep{Lattimer_Prakash_2007,Urbano_Veermae_2019} for the
maximum compactness, found by considering an affine fluid model
and imposing the hypothesis that such a model yields the most compact
star composed of a perfect fluid verifying a barotropic equation of
state. The analysis in this section, however, is independent of the
equation of state of the equilibrium star and follows from simply
considering the extreme case of the causality condition for the perturbed fluid. Consequently,
this upper bound for instability is universal
and it is not expected to be saturated
by any perfect fluid star solution verifying a physically meaningful
equation of state.

\section{Conclusion\label{sec:Conclusions}}

We have presented a system of first order differential equations to
describe general linear adiabatic, radial perturbations of spatially
compact, static, spherically symmetric solutions of general relativity
with a perfect fluid source. The new results do not rely on auxiliary
test functions nor on the introduction of a singular point at the
boundary of the star. Contrary to previous approaches, assuming some
regularity conditions for the equilibrium spacetime, the system can
be solved analytically, finding general solutions for the perturbation
variables.

The results were then used to study adiabatic, radial perturbations
of classical exact solutions of the Einstein field equations, computing
the first eigenfrequencies for
particular values of the spacetime parameters. 
For the considered models, the equilibria proved to be stable. We have also plotted the eigenfunctions associated with those eigenfrequencies, illustrating their radial profile. At first glance, the plots in Figures~\ref{fig:Eigenfunctions_Schwarzschild}--\ref{fig:Eigenfunctions_HeintIIa}  look remarkably similar.  However, this is a product of the very nature of the Sturm-Liouville problem, the constraints imposed by the boundary conditions, and the normalized radial coordinate we have used.

We have considered only a small subset of physically relevant exact solutions, leaving
aside some important solutions that have been used in the past to
study physically meaningful scenarios. As was discussed in Ref.~\cite{Luz_Carloni_2024a},
although the thermodynamical description of the perturbed matter fluid
is simpler in a comoving frame, this leads to extra complexity to
describe its dynamics such that we end up with a three-by-three system
of differential equations with a constraint, Eqs.~\eqref{eq:Comoving_Radial_Adiabatic_pdot_hat}--\eqref{eq:Comoving_Radial_Adiabatic_vsquare_constraint}.
Indeed, the formulation of the problem can be made more computationally
efficient. By changing the frame of reference, we can significantly simplify
the problem and efficiently study adiabatic, radial perturbations
of more complex background solutions. In other frames, however,  the stress-energy tensor
is non-diagonal, containing fluxes and anisotropic components. Therefore, 
great care has to be taken to ascertain what is meant by adiabatic
perturbations in the new frame. This will be done elsewhere.

The newly found system of equations also allowed us to conjecture, 
independently of the information on the equation of
state of the equilibrium fluid,
the upper bound $M/r_{b}\approx0.368$ for the maximum
compactness of a stable, static, self-gravitating perfect fluid.
The result relies on the hypothesis
that the interior Schwarzschild solution represents the extreme case
for a perfect fluid star. Hence, the maximum bound for the compactness
should be absolute, in the sense that no physically meaningful solution
of general relativity is expected to saturate it. Nonetheless, the
result is a tighter upper bound compared to the previously suggested
bounds in the literature, found by considering a specific barotropic
equation of state. We remark, however, that the 
reasoning for this result omits various important issues on the stability of compact
stellar objects. One example concerns the evolutionary timescales. The eigenfrequencies
$\left\{ \lambda_n\right\}_{n\in\mathbb{N}}$ can be used to determine the oscillation period or the
period of e-folds of each mode.
In the case of $\lambda_{0}\in\mathbb{R}$,
the star will oscillate, and the period of oscillation for the fundamental
mode is given by $T=2\pi/\left|\lambda_{0}\right|$. On the other
hand, if $\lambda_{0}$ is purely imaginary, the perturbations will not be bounded
and the quantity $T=2\pi/\left|i\lambda_{0}\right|$
represents the period of e-folds.
Since the fundamental mode
has the shortest growth timescale, we can assess the growth
timescale of the radial instabilities in the considered idealized
setup, considering this mode. Nonetheless, the evolutionary timescale
of compact stellar objects depends on various factors omitted
in our analysis, where we have assumed the idealized scenario of an
adiabatic perturbation. Indeed, the thermodynamic evolution of the
matter fields plays a predominant role in the evolution of those objects,
such that even if the star is pulsationally unstable, the radial modes
might not have time to grow and affect the structure. Moreover, 
the previous discussion does not consider the issue
of quasi-stability. In principle, a dynamically unstable object, but
such that its evolution would have a characteristic time comparable
to the timescale of the universe, might be effectively considered
stable. 
A realistic analysis of the topic should take into consideration all
these aspects. The presented upper bound for dynamical stability 
should then be valid in an idealized scenario, disregarding
any other effects.
\begin{acknowledgments}
PL acknowledges financial support provided under the European Union's
H2020 ERC Advanced Grant ``Black holes: gravitational engines of
discovery'' grant agreement no. Gravitas--101052587. Views and opinions
expressed are, however, those of the author only and do not necessarily
reflect those of the European Union or the European Research Council.
Neither the European Union nor the granting authority can be held
responsible for them. PL acknowledges support from the Villum Investigator
program supported by the VILLUM Foundation (grant no. VIL37766) and
the DNRF Chair program (grant no. DNRF162) by the Danish National
Research Foundation.

This project has received funding from the European Union's Horizon
2020 research and innovation programme under the Marie Sklodowska-Curie
grant agreement No 101007855 and No 101131233.

The work of SC has been carried out in the framework of activities
of the INFN Research Project QGSKY. 
\end{acknowledgments}

\appendix

\section{\label{sec:Appendix_1p1p2_decomposition}The 1+1+2 decomposition}

\subsection{Projectors and the Levi-Civita volume form}

Let $\left(\mathcal{M},g\right)$ be a Lorentzian manifold of dimension 4,
where $g$ represents the metric tensor, admitting in some open neighborhood
the existence of a congruence of timelike curves with tangent vector
field $u$. Let the congruence be affinely parameterized
and $u_{\alpha}u^{\alpha}=-1$. Without loss of generality, the manifold can be
foliated by 3-surfaces, $V$, pointwise orthogonal 
to the curves of the congruence, such that all tensorial quantities are
defined by their behavior along the direction of $u$ in $V$.
This formalism is called 1+3 spacetime covariant decomposition. This decomposition
of the spacetime manifold relies on the existence of a pointwise projector
to the cotangent space of $V$ to be defined as
\begin{equation}
h_{\alpha\beta}=g_{\alpha\beta}+u_{\alpha}u_{\beta}\,,\label{Def_eq:projector_h_definition}
\end{equation}
where $g_{\alpha\beta}$ represents the components of the metric tensor
in some local coordinate system. The projector operator verifies the following properties
\begin{equation}
\begin{aligned}h_{\alpha\beta} & =h_{\beta\alpha}\,, &  &  & h_{\alpha\beta}h^{\beta\gamma} & =h_{\alpha}{}^{\gamma}\,,\\
h_{\alpha\beta}u^{\alpha} & =0\,, &  &  & h_{\alpha}{}^{\alpha} & =3\,.
\end{aligned}
\end{equation}

The 1+1+2 decomposition builds from the 1+3 decomposition by defining
a spacelike congruence with tangent vector field $e$, such
that a tensorial quantity defined in the submanifold $V$ is characterized
by its behavior along $e$ and the 2-surfaces $W$, orthogonal to
both $u$ and $e$ at each point. Without loss of generality, let the spacelike
congruence to be affinely parameterized and $e_{\alpha}e^{\alpha}=1$.
We can define a projector onto the cotangent space of $W$ by
\begin{equation}
N_{\alpha\beta}=h_{\alpha\beta}-e_{\alpha}e_{\beta}\,,\label{Def_eq:projector_N_definition}
\end{equation}
verifying

\begin{equation}
\begin{aligned}N_{\alpha\beta} & =N_{\beta\alpha}\,, &  &  & N_{\alpha\beta}N^{\beta\gamma} & =N_{\alpha}{}^{\gamma}\,,\\
N_{\alpha\beta}u^{\alpha} & =N_{\alpha\beta}e^{\alpha}=0\,, &  &  & N_{\alpha}{}^{\alpha} & =2\,.
\end{aligned}
\label{Def_eq:projector_N_properties}
\end{equation}

It is also useful to introduce the following skew-symmetric
tensors derived from the covariant Levi-Civita tensor $\varepsilon_{\alpha\beta\gamma\sigma}$,
\begin{equation}
\begin{aligned}\varepsilon_{\alpha\beta\gamma} & =\varepsilon_{\alpha\beta\gamma\sigma}u^{\sigma}\,,\\
\varepsilon_{\alpha\beta} & =\varepsilon_{\alpha\beta\gamma}e^{\gamma}\,.
\end{aligned}
\label{Def_eq:volume_forms}
\end{equation}

In what follows, we will adopt the convention to indicate the symmetric and
anti-symmetric part of a tensor using parentheses and brackets, such
that for a 2-tensor $\chi$
\begin{equation}
\begin{aligned}\chi_{\left(\alpha\beta\right)} & =\frac{1}{2}\left(\chi_{\alpha\beta}+\chi_{\beta\alpha}\right)\,, &  &  & \chi_{\left[\alpha\beta\right]} & =\frac{1}{2}\left(\chi_{\alpha\beta}-\chi_{\beta\alpha}\right)\,.\end{aligned}
\end{equation}

\subsection{\label{Appendix:subsec_Covariant_derivatives_decomposition}Covariant
derivatives of $u$ and $e$}

Using the projector operators $h$ and $N$, the
covariant derivatives of the tangent vector fields $u$ and $e$ can
be uniquely decomposed at each point in their components along $u$,
$e$ and in $W$. This decomposition yields the so-called kinematical
quantities of the congruences formed by the integral curves of $u$
and $e$, providing a clear geometric and physical interpretation
of the behavior of the congruence.

Given a tensor quantity $\chi$, it is useful to introduce the compact 
notation
\begin{equation}
	\begin{aligned}
		\dot{\chi} & :=u^{\mu} \nabla_\mu \chi\,, &  &  & \widehat{\chi} & :=e^{\mu} \nabla_\mu \chi\,,
	\end{aligned}
\end{equation}
to represent the derivative along the integral curves of the vector field $u$ and the derivative along the integral curves of the vector field $e$, respectively.

In the 1+1+2 formalism, the covariant derivative of the tensor field
$u$ can be decomposed as
\begin{equation}
\nabla_{\alpha}u_{\beta}=-u_{\alpha}\left(\mathcal{A}e_{\beta}+\mathcal{A}_{\beta}\right)+\frac{1}{3}h_{\alpha\beta}\theta+\sigma_{\alpha\beta}+\omega_{\alpha\beta}\,,
\end{equation}
where
\begin{equation}
\begin{aligned}\mathcal{A} & =-u_{\mu}u^{\nu}\nabla_{\nu}e^{\mu}\,, &  &  & \mathcal{A}_{\alpha} & =N_{\alpha\mu}\dot{u}^{\mu}\,,\end{aligned}
\end{equation}
and the quantities $\theta$, $\sigma_{\alpha\beta}$ and $\omega_{\alpha\beta}$
are the kinematical quantities of the congruence of the integral
curves of $u$. Namely, $\theta$ is the expansion scalar, $\sigma_{\alpha\beta}$
is the shear tensor, and $\omega_{\alpha\beta}$ the vorticity tensor, defined as
\begin{equation}
\begin{aligned}\theta & =h^{\mu\nu}\nabla_{\mu}u_{\nu}\,,\\
\sigma_{\alpha\beta} & =\left(\frac{h_{\alpha}{}^{\mu}h_{\beta}{}^{\nu}+h_{\alpha}{}^{\nu}h_{\beta}{}^{\mu}}{2}-\frac{1}{3}h_{\alpha\beta}h^{\mu\nu}\right)\nabla_{\mu}u_{\nu}=\Sigma_{\alpha\beta}+2\Sigma_{(\alpha}e_{\beta)}+\Sigma\left(e_{\alpha}e_{\beta}-\frac{1}{2}N_{\alpha\beta}\right)\,,\\
\omega_{\alpha\beta} & =\frac{1}{2}\left(h_{\alpha}{}^{\mu}h_{\beta}{}^{\nu}-h_{\alpha}{}^{\nu}h_{\beta}{}^{\mu}\right)\nabla_{\mu}u_{\nu}=\varepsilon_{\alpha\beta\mu}\left(\Omega e^{\mu}+\Omega^{\mu}\right)\,,
\end{aligned}
\end{equation}
with
\begin{equation}
\begin{aligned}\Sigma_{\alpha\beta} & =\left(\frac{N_{\alpha}{}^{\mu}N_{\beta}{}^{\nu}+N_{\alpha}{}^{\nu}N_{\beta}{}^{\mu}}{2}-\frac{1}{2}N_{\alpha\beta}N^{\mu\nu}\right)\sigma_{\mu\nu}\,, &  &  & \Sigma_{\alpha} & =N_{\alpha}{}^{\mu}e^{\nu}\sigma_{\mu\nu}\,, &  &  & \Sigma & =e^{\mu}e^{\nu}\sigma_{\mu\nu}\,,\end{aligned}
\end{equation}
and
\begin{equation}
\begin{aligned}\Omega^{\alpha} & =\frac{1}{2}N_{\gamma}{}^{\alpha}\varepsilon^{\mu\nu\gamma}\nabla_{\mu}u_{\nu}\,, &  &  & \Omega & =\frac{1}{2}\varepsilon^{\mu\nu}\nabla_{\mu}u_{\nu}\,.\end{aligned}
\end{equation}
From their definitions, it is immediate to conclude that the covariantly defined vector and 2-tensor quantities
characterize the behavior of the kinematical quantities on the surfaces
$W$, whereas the scalars characterize the behavior along $u$ or
$e$.

Similarly, we can decompose the covariant derivative of $e$ along
$u$, $e$ and onto $W$ as
\begin{equation}
\nabla_{\alpha}e_{\beta}=\frac{1}{2}N_{\alpha\beta}\phi+\zeta_{\alpha\beta}+\varepsilon_{\alpha\beta}\xi+e_{\alpha}a_{\beta}-u_{\alpha}\alpha_{\beta}-\mathcal{A}u_{\alpha}u_{\beta}+\left(\frac{1}{3}\theta+\Sigma\right)e_{\alpha}u_{\beta}+\left(\Sigma_{\alpha}-\varepsilon_{\alpha\mu}\Omega^{\mu}\right)u_{\beta}\,,\label{Def_eq:Cov_dev_vector_e-1}
\end{equation}
where
\begin{equation}
\begin{aligned}\phi & =N^{\mu\nu}\nabla_{\mu}e_{\nu}\,, &  &  & \zeta_{\alpha\beta} & =\left(\frac{N_{\alpha}{}^{\mu}N_{\beta}{}^{\nu}+N_{\alpha}{}^{\nu}N_{\beta}{}^{\mu}}{2}-\frac{1}{2}N_{\alpha\beta}N^{\mu\nu}\right)\nabla_{\mu}e_{\nu}\,, &  &  & \xi & =\frac{1}{2}\varepsilon^{\mu\nu}\nabla_{\mu}e_{\nu}\,,\end{aligned}
\end{equation}
represent, respectively, the expansion scalar, the shear tensor, and
the twist of the congruence of the integral curves of $e$ projected
on $W$, and
\begin{equation}
\begin{aligned}a_{\alpha} & =e^{\mu}h_{\alpha}{}^{\nu}\nabla_{\mu}e_{\nu}\,, &  &  & \alpha_{\alpha} & =u^{\mu}h_{\alpha}{}^{\nu}\nabla_{\mu}e_{\nu}\,.\end{aligned}
\end{equation}

\subsection{Weyl and stress-energy tensors}

For the Levi-Civita connection, the Riemann tensor can be defined
by the Ricci identity for an arbitrary 1-form $\chi$:
\begin{equation}
R_{\alpha\beta\delta}{}^{\rho}\chi{}_{\rho}=\left(\nabla_{\alpha}\nabla_{\beta}-\nabla_{\beta}\nabla_{\alpha}\right)\chi_{\delta}\,.\label{Def_eq:Ricci_identity}
\end{equation}

In the case of a manifold of dimension four, the components of the
Riemann curvature tensor, $R_{\alpha\beta\gamma\delta}$, are given by
\begin{equation}
R_{\alpha\beta\gamma\delta}=C_{\alpha\beta\gamma\delta}+R_{\alpha\left[\gamma\right.}g_{\left.\delta\right]\beta}-R_{\beta\left[\gamma\right.}g_{\left.\delta\right]\alpha}-\frac{1}{3}R\,g_{\alpha\left[\gamma\right.}g_{\left.\delta\right]\beta}\,,\label{Def_eq:Weyl_tensor_definition}
\end{equation}
where $C_{\alpha\beta\gamma\delta}$ represents the Weyl tensor. The
Weyl tensor plays a crucial role in relativistic gravity, describing
the tidal forces, frame-dragging, and the properties of gravitational waves. Remarkably,
the Weyl 4-tensor can be completely characterized by two 2-tensors
\begin{align}
E_{\alpha\beta} & =C_{\alpha\mu\beta\nu}u^{\mu}u^{\nu}\,,\label{Def_eq:Weyl_tensor_electric}\\
H_{\alpha\beta} & =\frac{1}{2}\varepsilon_{\alpha}{}^{\mu\nu}C_{\mu\nu\beta\delta}u^{\delta}\,,
\end{align}
respectively referred as the ``electric'' and ``magnetic'' part
of the Weyl tensor, both symmetric and traceless tensors, such that
\begin{equation}
C_{\alpha\beta\gamma\delta}=-\varepsilon_{\alpha\beta\mu}\varepsilon_{\gamma\delta\nu}E^{\nu\mu}-2u_{\alpha}E_{\beta\left[\gamma\right.}u_{\left.\delta\right]}+2u_{\beta}E_{\alpha\left[\gamma\right.}u_{\left.\delta\right]}-2\varepsilon_{\alpha\beta\mu}H^{\mu}{}_{\left[\gamma\right.}u_{\left.\delta\right]}-2\varepsilon_{\mu\gamma\delta}H^{\mu}{}_{\left[\alpha\right.}u_{\left.\beta\right]}\,.\label{Def_eq:Weyl_tensor_1+3_decomposition}
\end{equation}

In the 1+1+2 covariant formalism, the components of
the Weyl tensor are themselves decomposed as
\begin{equation}
\begin{aligned}E_{\alpha\beta} & =\mathcal{E}\left(e_{\alpha}e_{\beta}-\frac{1}{2}N_{\alpha\beta}\right)+\mathcal{E}_{\alpha}e_{\beta}+e_{\alpha}\mathcal{E}_{\beta}+\mathcal{E}_{\alpha\beta}\,,\\
H_{\alpha\beta} & =\mathcal{H}\left(e_{\alpha}e_{\beta}-\frac{1}{2}N_{\alpha\beta}\right)+\mathcal{H}_{\alpha}e_{\beta}+e_{\alpha}\mathcal{H}_{\beta}+\mathcal{H}_{\alpha\beta}\,,
\end{aligned}
\end{equation}
where
\begin{equation}
\begin{aligned}\mathcal{E} & =E_{\mu\nu}e^{\mu}e^{\nu}=-N^{\mu\nu}E_{\mu\nu}\,, &  &  & \mathcal{H} & =e^{\mu}e^{\nu}H_{\mu\nu}=-N^{\mu\nu}H_{\mu\nu}\,,\\
\mathcal{E}_{\alpha} & =N_{\alpha}{}^{\mu}e^{\nu}E_{\mu\nu}=e^{\mu}N_{\alpha}{}^{\nu}E_{\mu\nu}\,, &  &  & \mathcal{H}_{\alpha} & =N_{\alpha}{}^{\mu}e^{\nu}H_{\mu\nu}=e^{\mu}N_{\alpha}{}^{\nu}H_{\mu\nu}\,,\\
\mathcal{E}_{\alpha\beta} & =E_{\left\{ \alpha\beta\right\} }\,, &  &  & \mathcal{H}_{\alpha\beta} & =H_{\left\{ \alpha\beta\right\} }\,.
\end{aligned}
\label{Def_eq:Weyl_tensor_components_definition}
\end{equation}

Lastly, to write the Einstein field equations in the language of the
1+1+2 formalism, we need the covariant decomposition of the metric stress-energy tensor $T_{\alpha\beta}$.
Then,
\begin{equation}
T_{\alpha\beta}=\mu\,u_{\alpha}u_{\beta}+\left(p+\Pi\right)e_{\alpha}e_{\beta}+\left(p-\frac{1}{2}\Pi\right)N_{\alpha\beta}+2Qe_{\left(\alpha\right.}u_{\left.\beta\right)}+2Q_{\left(\alpha\right.}u_{\left.\beta\right)}+2\Pi_{\left(\alpha\right.}e_{\left.\beta\right)}+\Pi_{\alpha\beta}\,,\label{Def_eq:Stress-energy_tensor_decomposition}
\end{equation}
with 
\begin{equation}
\begin{aligned}\mu & =u^{\mu}u^{\nu}T_{\mu\nu}\,, &  &  & Q_{\alpha} & =-N_{\alpha}{}^{\mu}u^{\nu}T_{\mu\nu}\,,\\
p & =\frac{1}{3}\left(e^{\mu}e^{\nu}+N^{\mu\nu}\right)T_{\mu\nu}\,, &  &  & \Pi_{\alpha} & =N_{\alpha}{}^{\mu}e^{\nu}T_{\mu\nu}\,,\\
\Pi & =\frac{1}{3}\left(2e^{\mu}e^{\nu}-N^{\mu\nu}\right)T_{\mu\nu}\,, &  &  & \Pi_{\alpha\beta} & =\left(\frac{N_{\alpha}{}^{\mu}N_{\beta}{}^{\nu}+N_{\alpha}{}^{\nu}N_{\beta}{}^{\mu}}{2}-\frac{1}{2}N_{\alpha\beta}N^{\mu\nu}\right)T_{\mu\nu}\,.\\
Q & =-e^{\mu}u^{\nu}T_{\mu\nu}\,,
\end{aligned}
\label{Def_eq:Energy_momentum_tensor_decomposition_quantities}
\end{equation}
The various contributions in the covariant decomposition of the stress-energy
tensor in Eq.~(\ref{Def_eq:Energy_momentum_tensor_decomposition_quantities})
have direct physical meaning. Given an observer with 4-velocity $u$,
$\mu$ represents the mass-energy density of the fluid,
$p$ the isotropic pressure, $Q$ and $Q_{\alpha}$ represent, respectively, heat
and momentum flows along $e$ and in $W$, and $\Pi$,
$\Pi_{\alpha}$ and $\Pi_{\alpha\beta}$ characterize the anisotropic
pressure within the fluid.

\section{\label{sec:Appendix_Scalar_perturbation_equations}1+1+2 scalar perturbation
equations}

Using the 1+1+2 formalism, briefly introduced in Appendix~\ref{sec:Appendix_1p1p2_decomposition}, we list here the covariant version of the equations that describe the scalar
adiabatic perturbations of a fluid distribution in a static, LRS II background. These equations are written in terms of the variables in Eq.~\eqref{eq:GI_dot_derivatives_definition}
and characterize the perturbation from the point of view of an observer locally
comoving with the matter composing the relativistic compact stellar
object. They read:
\begin{equation}
\begin{aligned}\widehat{\mathsf{A}}-\ddot{\theta} & =\frac{1}{2}\left(\mathsf{m}+3\mathsf{p}\right)+\widehat{\mathcal{A}}_{0}\left(\frac{1}{3}\theta+\Sigma\right)-\left(3\mathcal{A}_{0}+\phi_{0}\right)\mathsf{A}-\mathcal{A}_{0}\mathsf{F}\,,\end{aligned}
\label{eq_Appendix:Raychaudhuri_eq}
\end{equation}
\begin{equation}
\begin{aligned}\widehat{\mathsf{p}} & =\left(\frac{1}{3}\theta+\Sigma\right)\widehat{p}_{0}-\mathcal{A}_{0}\left(\mathsf{m}+2\mathsf{p}\right)-\left(\mu_{0}+p_{0}\right)\mathsf{A}\,,\end{aligned}
\label{eq_Appendix:pert_momentum_cons}
\end{equation}
\begin{equation}
\ddot{\Sigma}-\frac{2}{3}\ddot{\theta}=\frac{1}{3}\left(\mathsf{m}+3\mathsf{p}\right)-\mathsf{E}-\mathcal{A}_{0}\mathsf{F}-\phi_{0}\mathsf{A}\,,
\end{equation}
\begin{equation}
\frac{2}{3}\widehat{\theta}-\widehat{\Sigma}=\frac{3}{2}\phi_{0}\Sigma\,.\label{Appendix:Theta_hat_Sigma_hat}
\end{equation}
\begin{equation}
\begin{aligned}\mathsf{E} & =\mathcal{E}_{0}\left(\frac{3}{2}\Sigma-\theta\right)-\frac{1}{2}\left(\mu_{0}+p_{0}\right)\Sigma\,,\end{aligned}
\label{eq_Appendix:dotE_eq}
\end{equation}
\begin{equation}
\mathsf{F}=\left(2\mathcal{A}_{0}-\phi_{0}\right)\left(\frac{1}{3}\theta-\frac{1}{2}\Sigma\right)\,,\label{eq_Appendix:dot_phi_eq}
\end{equation}
\begin{equation}
\mathsf{m}=-\left(\mu_{0}+p_{0}\right)\theta\,,
\end{equation}

Upon reorganization, using the background gravitational field
equations and applying a harmonic decomposition of the perturbation
variables, the above equations can be recast into the system in Sec.~\ref{Pert_Eq.}.

The perturbation equations become increasingly more complicated as
we relax the constraints that we have assumed before. The general
set of perturbation equations for the variables in Eq.~\eqref{eq:GI_dot_derivatives_definition},
together with additional perturbation variables useful in other specific
frames can be found in~\cite{Luz_Carloni_2024a}.

\section{\label{sec:Appendix_Pulsation_equation} Derivation of Chandrasekhar's radial pulsation equation}

In this Appendix, we show
how the system of equations for the comoving observer in Appendix~\ref{sec:Appendix_Scalar_perturbation_equations}
returns the well-known Chandrasekhar radial pulsation equation~\citep{Chandrasekhar_1964_PRL,Chandrasekhar_1964_ApJ}.
We will then start by recalling the classical derivation of the Chandrasekhar equation. Then, we will break covariance and gauge invariance in the 1+1+2 perturbation equations to prove that they lead to the same result.

Let us start by constructing the Chandrasekhar pulsation equation considering perturbations of the metric directly. Consider the Schwarzschild coordinate system $\left(t,r,\psi,\varphi\right)$
defined by an observer at spatial infinity, such that events in both
the equilibrium and the perturbed spacetime can be identified unambiguously. Let the equilibrium
spacetime be characterized by a line element of the form
\begin{equation}
ds^{2}=(g_0)_{\alpha\beta}dx^\alpha dx^\beta=-e^{2\Phi_{0}(r)}dt^{2}+e^{2\Lambda_{0}(r)}dr^{2}+r^{2}d\Omega^{2}\,,\label{Bckg_g}
\end{equation}
where $d\Omega^{2}=d\psi^{2}+\sin\psi d\varphi^{2}$ represents the
natural line element of the unit 2-sphere. We will adopt the nomenclature of the body of the text and indicate
quantities in the background spacetime by a subscript ``0''.

Assuming the equilibrium spacetime to be permeated by a perfect fluid, such that, from the point of view of an observer at rest with matter, the energy-momentum tensor will be written as in Eq.~\eqref{T0Ord}
\begin{equation}
T_{\alpha\beta}=\left(\mu_{0}+p_{0}\right)(u_{0})_{\alpha}(u_{0})_{\beta}+p_{0}\left(g_{0}\right)_{\alpha\beta}\,,
\end{equation}
where
\begin{equation}
\left(u_{0}\right)_{\alpha}dx^{\alpha}  =-e^{\Phi_{0}}dt\,,
\label{pert_vec_u_0}
\end{equation}
the Einstein field equations are
\begin{align}
\Lambda'_{0} & =\frac{1}{2}e^{2\Lambda_{0}}\mu_{0}r-\frac{e^{2\Lambda_{0}}}{2r}+\frac{1}{2r}\,,\\
\Phi'_{0} & =\frac{1}{2}e^{2\Lambda_{0}}p_{0}r+\frac{e^{2\Lambda_{0}}}{2r}-\frac{1}{2r}\,,\\
\Phi''_{0} & =\frac{1}{4}e^{2\Lambda_{0}}\mu_{0}+\frac{1}{4}e^{4\Lambda_{0}}\mu_{0}+\frac{5}{4}e^{2\Lambda_{0}}p_{0}-\frac{3}{4}e^{4\Lambda_{0}}p_{0}+\frac{1}{4}e^{4\Lambda_{0}}\mu_{0}p_{0}r^{2}-\frac{1}{4}e^{4\Lambda_{0}}p_{0}^{2}r^{2}-\frac{e^{4\Lambda_{0}}}{2r^{2}}+\frac{1}{2r^{2}}\,,
\end{align}
and the Bianchi identity
\begin{equation}
p'_{0}=-\Phi_{0}\left(\mu_{0}+p_{0}\right)\,.
\end{equation}

Now, radially perturbing the equilibrium configuration leads to a new metric associated with the line element 
\begin{equation}
ds^{2}=g_{\alpha\beta}dx^\alpha dx^\beta=-e^{2\Phi(t,r)}dt^{2}+e^{2\Lambda(t,r)}dr^{2}+r^{2}d\Omega^{2}\,,
\label{Pert_g}
\end{equation}
and the energy-momentum tensor for an observer comoving with matter will be described by
\begin{equation}
T_{\alpha\beta}=\left(\mu+p\right)u_{\alpha}u_{\beta}+p g_{\alpha\beta}\,,
\end{equation}
where $u_\alpha$ represent, in a local coordinate system, the components of the 4-velocity form of an observer comoving with the perturbed fluid.
The perturbation will induce a change in a given scalar variable $X$, which can be characterized by the quantity
\begin{equation}
\delta X=X\left(t,r\right)-X_{0}\left(r\right)\,.
\end{equation}
This expression constitutes the Eulerian representation of perturbations, i.e., the description of
the perturbation from the point of view of an observer in a 
position with constant $r$ coordinate.

In addition to the $\delta$-variations, we need to introduce an extra parameter to describe the radial displacement of a fluid element. A fluid element at the coordinate $r$ in the unperturbed spacetime,  is moved to a new position $\tilde{r}$, in the perturbed spacetime, such that, 
\begin{equation}
	\tilde{r}=r+\eta\left(t,r\right)\,.
\end{equation}
The parameter $\eta$ is usually called ``radial displacement''.
We need to be especially careful in understanding the meaning of $\eta$.  Even if this function is connected to the choice of a local coordinate system,  $\eta$ effectively also characterizes the mapping between the unperturbed and perturbed spacetimes, as it is relabeling events. Therefore, it cannot be just considered a product of a coordinate transformation, but rather a gauge. For this reason, $\eta$ is often called  ``gauge parameter''.
In this appendix, for example, in our construction, we have tacitly considered the Schwarzschild coordinate system to write the components of the perturbed metric and those of the background metric. Therefore, we will find that the perturbations depend directly on the gauge parameter. This, of course, is not a necessity, but it is useful in the treatment of this type of perturbations.
We stress, however, that this approach should be used with
caution.  In the case of adiabatic,
radial perturbations, it is indeed possible to define the same coordinate
system in both the equilibrium and the perturbed spacetimes. Therefore,
the conclusions are not ambiguous. However, in
general this is not the case, and attributing physical observable
effects to active coordinate transformations, might lead to gauge-dependent,
physically meaningless conclusions.  In addition, choosing from the start
the coordinate system of the equilibrium and the perturbed spacetimes will force the analysis
to a specific gauge, which might lead to spurious complexity of the equations.

In terms of Eulerian perturbations, at first  perturbative order, that is, disregarding quadratic and higher-order terms of
the $\delta$-variations, we have 
\begin{equation}
u_{\alpha}dx^{\alpha}  =-e^{\Phi_{0}}\left(1+\delta\Phi\right)dt+e^{2\Lambda_{0}-\Phi_{0}}\eta\eta^{\ast}dr\,.\\ \label{pert_vec_u}
\end{equation}
where we have indicated the derivative with respect to $t$ with an ``asterisk''. Then the perturbed gravitational field equations  associated with the metric Eq.~\eqref{Pert_g} read, 
\begin{align}
\left(\delta\Lambda\right){}^{\ast}= & -\left(\Lambda'_{0}+\Phi'_{0}\right)\eta^{\ast}\,,\label{EFE01}\\
\left(\delta\Lambda\right)'= & \left(2r\Lambda'_{0}-\frac{1}{r}\right)\delta\Lambda+\frac{1}{2}re^{2\Lambda_{0}}\delta\mu\,,\\
\left(\delta\Phi\right)'= & \left(2\Phi'_{0}+\frac{1}{r}\right)\delta\Lambda+\frac{1}{2}re^{2\Lambda_{0}}\delta p\,,\label{EFE11}\\
\left(\delta\Phi\right)''= & e^{2(\Lambda_{0}-\Phi_{0})}\delta\Lambda^{\ast\ast}+\frac{1}{2}e^{2\Lambda_{0}}\left(1+r\Phi'_{0}\right)\delta\mu+\frac{1}{2}e^{4\Lambda_{0}}\left(1+r\Lambda'_{0}-2r\Phi'_{0}\right)\delta p\nonumber \\
 & -\frac{e^{2\Lambda_{0}}}{r^{2}}\left[2e^{2\Lambda_{0}}-r\Lambda'_{0}\left(3+4r\Phi'_{0}\right)+r\Phi'_{0}\left(1+4r\Phi'_{0}\right)\right]\delta\Lambda\,,
\end{align}
and the Bianchi identities
\begin{align}
 & \delta\mu^{\ast}=\frac{1}{r}\left(-2\eta^{\ast}+r\eta^{\ast}\Phi'_{0}-r\eta'^{\ast}\right)\left(\mu_{0}+p_{0}\right)-\eta^{\ast}\mu'_{0}\,,\label{BI_mu}\\
 & \left(\delta p\right)'+\eta^{\ast\ast}e^{2(\Lambda_{0}-\Phi_{0})}\left(\mu_{0}+p_{0}\right)+\left(\delta\Phi\right)'\left(\mu_{0}+p_{0}\right)+\left(\delta\mu+\delta p\right)\Phi'_{0}=0\,,\label{BI_p}
\end{align}
where we have indicated the derivative with respect to $r$ with a
``prime''.
Integrating Eq.~\eqref{BI_mu} with respect to $t$ leads to 
\begin{equation}
\delta\mu=\frac{1}{r}\left(-2\eta+r\eta\Phi'_{0}-r\eta'\right)\left(\mu_{0}+p_{0}\right)-\eta\mu'_{0}\,,\label{BI_mu_integrated}
\end{equation}
where we have set the integration constant to zero as we assume $\delta\mu=0$
when $\eta=0$.
As we have mentioned 
before, this choice of integration constant follows from encoding the perturbations  
in the gauge
parameter. Consistently, in the above expressions, if $\eta=0$, we recover the 
background spacetime.

Continuing, the system above is closed by providing an equation of
state that relates the pressure to the energy density. Following Chandrasekhar's
original derivation, we will consider a barotropic equation of state,
such that $p=f\left(\mu\right)$. Using the equation of state, Eq.~\eqref{BI_mu_integrated}
allows us to write $\delta p$ in terms of $\eta$ and its derivatives:
\begin{equation}
\delta p=\frac{\Gamma_{1}p_{0}}{r}\left(-2\eta+r\eta\Phi'_{0}-r\eta'\right)-\eta p'_{0}\,,
\end{equation}
where the quantity
\begin{equation}
\Gamma_{1}=\frac{\mu_{0}+p_{0}}{p_{0}}f'\left(\mu_{0}\right)\,,
\end{equation}
is the adiabatic index of the fluid. Integrating Eq.~\eqref{EFE01}
with respect to $t$, choosing the integration constant such that
$\delta\Lambda=0$ when $\eta=0$, yields
\begin{equation}
\delta\Lambda=-\left(\Lambda'_{0}+\Phi'_{0}\right)\eta\,.\label{eq_Appendix:deltaLambda_integrated}
\end{equation}

Substituting the above relations and Eq.~\eqref{EFE11} in the momentum
conservation equation~\eqref{BI_p}, after a laborious yet straightforward
simplification, we obtain an equation for $\eta$, the so-called Chandrasekhar
radial pulsation equation. 

Alternatively to the Eulerian description, we can define a Lagrangian description of the perturbations. In this picture, an observer located at some point with radial coordinate
$r$ in the background spacetime is moved together with matter to $\tilde{r}$
in the perturbed spacetime.
We then define a Lagrangian perturbation of a
scalar quantity $X$ as 
\begin{equation}
\Delta X=X\left(t,\tilde{r}\right)-X_{0}\left(r\right)\,.
\end{equation}
The 1+1+2 perturbation variables
we have used in this paper, correspond, by definition, to Lagrangian perturbation
variables. Indeed, since all the quantities in Sec.~\ref{Pert_Eq.}
are defined from the point of view of a comoving observer, they are
naturally Lagrangian.

Thus, in order to prove that the equations
following from the covariant approach imply the Chandrasekhar equation,
we have to break covariance and gauge invariance and show that
the perturbation equations~\eqref{eq_Appendix:Raychaudhuri_eq}--\eqref{Appendix:Theta_hat_Sigma_hat}
reduce to the perturbed field equations or to the Bianchi identities~\eqref{EFE01}--\eqref{BI_p}.
In that regard, we have to relate the Lagrangian variables
$\left\{ \mathcal{A},\phi, \theta, \Sigma\right\} $ and $\left\{ \mathsf{m},\mathsf{p},\mathsf{A},\mathsf{F},\mathsf{E},\theta,\Sigma\right\} $
in terms of Eulerian variables to linear perturbation order. 

We already know how to write, at first order, the 1-form associated
with the vector field $u$  in terms of the metric coefficients
of the equilibrium and the perturbed spacetime, Eq.~\eqref{pert_vec_u}. For the $e$
congruence, we have
\begin{equation}
\begin{aligned}
\left(e_{0}\right)_{\alpha}dx^{\alpha} & =e^{\Lambda_{0}}dr\,,\\
e_{\alpha}dx^{\alpha} & =-e^{\Lambda_{0}}\eta^{\ast}dt+e^{\Lambda_{0}}\left(1+\delta\Lambda\right)dr\,,
\end{aligned}\label{pert_vec_e}
\end{equation}
Then, from the definitions in Appendix~\ref{sec:Appendix_1p1p2_decomposition} and Eqs.~\eqref{pert_vec_u_0}, \eqref{pert_vec_u}, \eqref{pert_vec_e}
we obtain 
\begin{align}
\mathcal{A}_0= & e^{-\Lambda_{0}}\Phi'_{0}\,,\\
\phi_0= & \frac{2e^{-\Lambda_{0}}}{r}\,,\\
\theta_0= & 0\,,\\
\Sigma_0= & 0\,,
\end{align}
 and
\begin{align}
\mathcal{A}= & e^{-\Lambda_{0}}\Phi'_{0}+e^{-\Lambda_{0}}\delta\Phi'-e^{-\Lambda_{0}}\Phi'_{0}\delta\Lambda+\eta^{\ast\ast}e^{\Lambda_{0}-2\Phi_{0}}\,,\\
\phi= & \frac{2e^{-\Lambda_{0}}}{r}\left(1-\delta\Lambda\right)\,,\\
\theta= & \frac{e^{-\Phi_{0}}}{r}\left(r\,\delta\Lambda^*+2\eta^{\ast}+r\eta^{\ast}\Lambda'_{0}+r\eta'^{\ast}\right)\,,\\
\Sigma= & \frac{2}{3}\frac{e^{-\Phi_{0}}}{r}\left(r\,\delta\Lambda^*+\eta^{\ast}+r\eta^{\ast}\Lambda'_{0}+r\eta'^{\ast}\right)\,.
\end{align}
Now, the following relation holds between the Lagrangian and Eulerian
perturbations of a scalar quantity $X$:
\begin{equation}
\Delta X=\delta X-X'_{0}\eta\,.
\end{equation}
This equation can be used to relate the Lagrangian variables $\left\{ \mathsf{m},\mathsf{p},\mathsf{A}\right\} $
with their Eulerian counterparts. For the Lagrangian perturbation
of the time derivative of the pressure, $\mathsf{p}$, we find
\begin{equation}
\mathsf{p}=\dot{p}-\mathring{p}_{0}\,,\label{PertTransf_p}
\end{equation}
where we have defined
\begin{equation}
\mathring{p}=\left(u_{0}\right)^{\alpha}\partial_{\alpha}p\,.
\end{equation}
Expressing the right-hand side of Eq.~\eqref{PertTransf_p} in terms
of Eulerian perturbations yields
\begin{equation}
\mathsf{p}=\left(\delta p\right)^{\boldsymbol\cdot}+p'_{0}\dot{\eta}=e^{-\Phi_{0}}\left[\left(\delta p\right)^{\ast}+p'_{0}\eta^{\ast}\right]\,.
\end{equation}
It is also useful to find the expression for the hat derivative of
$\mathsf{p}$ in terms of the metric perturbations:
\begin{equation}
\widehat{\mathsf{p}}=e^{-\Lambda_{0}}\Phi'_{0}\mathsf{p}-e^{-\left(\Phi_{0}+\Lambda_{0}\right)}\left[\left(\delta p\right)'^{\ast}+p''_{0}\eta^{\ast}+p'_{0}\eta'^{\ast}\right]\,.
\end{equation}
Similarly, we obtain the following expressions for the $\mathsf{m}$
and $\mathsf{A}$ quantities:
\begin{align}
\mathsf{m} & =e^{-\Phi_{0}}\left[\left(\delta\mu\right)^{\ast}+\mu'_{0}\eta^{\ast}\right]\,,\\
\mathsf{A} & =-e^{-\Lambda_{0}-3\Phi_{0}}\left[e^{2\Phi_{0}}\Phi'_{0}\delta\Lambda^{\ast}-e^{2\Phi_{0}}\left(\delta\Phi\right)'^{\ast}+e^{2\Phi_{0}}\left(\Lambda'_{0}\Phi'_{0}-\Phi''_{0}\right)\eta^{\ast}-e^{2\Lambda_{0}}\eta^{\ast\ast\ast}\right]\,.
\end{align}
Finally, using the above results, Eqs.~\eqref{eq:Background_phi_A_E},
\eqref{eq_Appendix:dotE_eq} and \eqref{eq_Appendix:dot_phi_eq}, we can obtain algebraically the  expressions for $\mathcal{E}_{0}$, $\mathsf{F}$ and $\mathsf{E}$ in terms of the coefficients of the metric in Eq.~\eqref{Pert_g} and their perturbations.

Gathering all the above results and substituting them in the perturbed
momentum conservation equation for the comoving frame, Eq.~\eqref{eq_Appendix:pert_momentum_cons},
yields, after some fairly long calculations, the momentum conservation
Eq.~\eqref{BI_p}. Applying the same substitutions in the remaining
equations of Appendix~\ref{sec:Appendix_Scalar_perturbation_equations}
yield a system of equations equivalent to combinations of the perturbed
Einstein equations~\eqref{EFE01}--\eqref{eq_Appendix:deltaLambda_integrated}. Thus, we conclude 
that in the comoving frame, the covariant, gauge-invariant
equations for adiabatic, radial perturbations  are
equivalent to the ones derived from the metric, coordinate based approach.

\section{\label{sec:Appendix_Matrix_A}Matrix $\mathds{A}$}

Consider the matrix $\Theta$ defined in Eq.~\eqref{eq:Radial_Adiabatic_comoving_Theta_matrix}.
To shorten the expressions, let $\Theta_{ij}$ represent the $ij$-entry
of $\Theta$, and $\left(\Theta_{n}\right)_{ij}$ be interpreted as
the $ij$-entry of the $n$th order coefficient of the power expansion
of $\Theta$ at $r=0$, Eq.~\eqref{eq:Theta_matrix_power_series}.
Then, the matrix $\mathds{A}$ in Eq.~\eqref{eq:Radial_Adiabatic_comoving_recurrence_relation}
is given by
\[
\mathds{A}=\left[\begin{array}{lcr}
A_{11} & A_{12} & A_{13}\\
A_{21} & A_{22} & A_{23}\\
A_{31} & A_{32} & A_{33}
\end{array}\right]\,,
\]
where
\begin{equation}
A_{11}=\Theta_{11}-\frac{1}{3}r^{2}\Theta_{31}\left[\left(\Theta_{0}\right)_{12}\left(\Theta_{0}\right)_{23}-3\left(\Theta_{1}\right)_{13}\right]\,,
\end{equation}
\begin{equation}
\begin{aligned}A_{12}= & -\frac{36\Theta_{13}}{r^{3}}+\frac{12}{r^{2}}\left[\Theta_{12}\left(\Theta_{0}\right)_{23}-\left(\Theta_{0}\right)_{12}\left(\Theta_{0}\right)_{23}+3\left(\Theta_{1}\right)_{13}\right]+\frac{12}{r}\left(\Theta_{33}-\Theta_{11}\right)\left[\left(\Theta_{0}\right)_{12}\left(\Theta_{0}\right)_{23}-3\left(\Theta_{1}\right)_{13}\right]\\
 & +\frac{2}{3}r^{2}\Theta_{32}\left[\left(\Theta_{0}\right)_{12}\left(\Theta_{0}\right)_{23}-3\left(\Theta_{1}\right)_{13}\right]\left\{ 6\left(\Theta_{0}\right)_{23}\left[\left(\Theta_{1}\right)_{22}-\left(\Theta_{1}\right)_{33}\right]-9\left(\Theta_{2}\right)_{23}+2\left(\Theta_{0}\right)_{32}\left[\left(\Theta_{0}\right)_{23}\right]^{2}\right\} \\
 & -4\Theta_{32}\left(\Theta_{0}\right)_{23}\left[\left(\Theta_{0}\right)_{12}\left(\Theta_{0}\right)_{23}-3\left(\Theta_{1}\right)_{13}\right]-12\Theta_{12}\left(\Theta_{0}\right)_{23}\left[\left(\Theta_{1}\right)_{22}-\left(\Theta_{1}\right)_{33}\right]+18\Theta_{12}\left(\Theta_{2}\right)_{23}\\
 & -4\Theta_{12}\left(\Theta_{0}\right)_{32}\left[\left(\Theta_{0}\right)_{23}\right]^{2}+4r\Theta_{31}\left[\left(\Theta_{0}\right)_{12}\left(\Theta_{0}\right)_{23}-3\left(\Theta_{1}\right)_{13}\right]^{2}\,,
\end{aligned}
\end{equation}

\begin{equation}
A_{13}=\frac{1}{3}r^{3}\Theta_{32}\left[\left(\Theta_{0}\right)_{12}\left(\Theta_{0}\right)_{23}-3\left(\Theta_{1}\right)_{13}\right]-r\Theta_{12}\,,
\end{equation}

\begin{equation}
A_{21}=-\frac{1}{36}r^{3}\Theta_{31}\,,
\end{equation}
\begin{equation}
\begin{aligned}A_{22} & =\frac{1}{9}r^{3}\Theta_{32}\left\{ 3\left(\Theta_{0}\right)_{23}\left[\left(\Theta_{1}\right)_{22}-\left(\Theta_{1}\right)_{33}\right]-\frac{9}{2}\left(\Theta_{2}\right)_{23}+\left[\left(\Theta_{0}\right)_{23}\right]^{2}\left(\Theta_{0}\right)_{32}\right\} \\
 & +\frac{1}{3}r^{2}\Theta_{31}\left[\left(\Theta_{0}\right)_{12}\left(\Theta_{0}\right)_{23}-3\left(\Theta_{1}\right)_{13}\right]-\frac{r}{3}\left(\Theta_{0}\right)_{23}\Theta_{32}+\Theta_{33}\,,
\end{aligned}
\end{equation}

\begin{equation}
A_{23}=\frac{1}{36}r^{4}\Theta_{32}\,,
\end{equation}

\begin{align}
A_{31} & =\frac{1}{9}\Theta_{31}\left\{ 3\left(\Theta_{0}\right)_{23}\left[r^{2}\left(\Theta_{1}\right)_{22}-r^{2}\left(\Theta_{1}\right)_{33}-1\right]-\frac{9}{2}r^{2}\left(\Theta_{2}\right)_{23}+r^{2}\left(\Theta_{0}\right)_{32}\left[\left(\Theta_{0}\right)_{23}\right]^{2}\right\} -\frac{\Theta_{21}}{r}\,,
\end{align}
\begin{equation}
\begin{aligned}A_{32}= & -\frac{2r}{3}\Theta_{31}\left[\left(\Theta_{0}\right)_{12}\left(\Theta_{0}\right)_{23}-3\left(\Theta_{1}\right)_{13}\right]\left\{ 6\left(\Theta_{0}\right)_{23}\left[\left(\Theta_{1}\right)_{22}-\left(\Theta_{1}\right)_{33}\right]-9\left(\Theta_{2}\right)_{23}+2\left[\left(\Theta_{0}\right)_{23}\right]^{2}\left(\Theta_{0}\right)_{32}\right\} \\
 & -r^{2}\Theta_{32}\left(\Theta_{0}\right)_{23}\left[\left(\Theta_{1}\right)_{22}-\left(\Theta_{1}\right)_{33}\right]\left\{ 4\left(\Theta_{0}\right)_{23}\left[\left(\Theta_{1}\right)_{22}-\left(\Theta_{1}\right)_{33}\right]-6\left(\Theta_{2}\right)_{23}+\frac{4}{3}\left[\left(\Theta_{0}\right)_{23}\right]^{2}\left(\Theta_{0}\right)_{32}\right\} \\
 & -r^{2}\Theta_{32}\left[\left(\Theta_{0}\right)_{23}\right]^{2}\left(\Theta_{0}\right)_{32}\left\{ \frac{4}{3}\left(\Theta_{0}\right)_{23}\left[\left(\Theta_{1}\right)_{22}-\left(\Theta_{1}\right)_{33}\right]-2\left(\Theta_{2}\right)_{23}+\frac{4}{9}\left[\left(\Theta_{0}\right)_{23}\right]^{2}\left(\Theta_{0}\right)_{32}\right\} \\
 & +\frac{36}{r^{4}}\left[\Theta_{23}-\left(\Theta_{0}\right)_{23}\right]-\frac{12}{r^{3}}\left(\Theta_{22}-\Theta_{33}\right)\left(\Theta_{0}\right)_{23}+\frac{12}{r^{2}}\Theta_{21}\left[\left(\Theta_{0}\right)_{12}\left(\Theta_{0}\right)_{23}-3\left(\Theta_{1}\right)_{13}\right]\\
 & +\frac{2}{r}\left(\Theta_{22}-\Theta_{33}\right)\left\{ 6\left(\Theta_{0}\right)_{23}\left[\left(\Theta_{1}\right)_{22}-\left(\Theta_{1}\right)_{33}\right]-9\left(\Theta_{2}\right)_{23}+2\left[\left(\Theta_{0}\right)_{23}\right]^{2}\left(\Theta_{0}\right)_{32}\right\} \\
 & +\frac{4}{3}\left(\Theta_{0}\right)_{23}\Theta_{32}\left\{ 6\left(\Theta_{0}\right)_{23}\left[\left(\Theta_{1}\right)_{22}-\left(\Theta_{1}\right)_{33}\right]-9\left(\Theta_{2}\right)_{23}+2\left[\left(\Theta_{0}\right)_{23}\right]^{2}\left(\Theta_{0}\right)_{32}\right\} \\
 & +r^{2}\Theta_{32}\left(\Theta_{2}\right)_{23}\left\{ 6\left(\Theta_{0}\right)_{23}\left[\left(\Theta_{1}\right)_{22}-\left(\Theta_{1}\right)_{33}\right]-9\left(\Theta_{2}\right)_{23}+2\left[\left(\Theta_{0}\right)_{23}\right]^{2}\left(\Theta_{0}\right)_{32}\right\} \\
 & +\frac{2}{r^{2}}\left\{ 6\left(\Theta_{0}\right)_{23}\left[\left(\Theta_{1}\right)_{22}-\left(\Theta_{1}\right)_{33}\right]-9\left(\Theta_{2}\right)_{23}-2\left[\left(\Theta_{0}\right)_{23}\right]^{2}\left[\Theta_{32}-\left(\Theta_{0}\right)_{32}\right]\right\} \\
 & +\frac{4}{r}\Theta_{31}\left(\Theta_{0}\right)_{23}\left[\left(\Theta_{0}\right)_{12}\left(\Theta_{0}\right)_{23}-3\left(\Theta_{1}\right)_{13}\right]\,,
\end{aligned}
\end{equation}

\begin{equation}
A_{33}=\frac{r}{9}\Theta_{32}\left\{ 3\left(\Theta_{0}\right)_{23}\left(1-r^{2}\left[\left(\Theta_{1}\right)_{22}-\left(\Theta_{1}\right)_{33}\right]\right)+\frac{9}{2}r^{2}\left(\Theta_{2}\right)_{23}-\left[\left(\Theta_{0}\right)_{23}\right]^{2}\left(\Theta_{0}\right)_{32}r^{2}\right\} +\Theta_{22}\,.
\end{equation}

At first glance, it might seem that the $\mathds{A}$ matrix is singular at $r=0$. However, after direct substitution and simplification, one can verify that $\mathds{A}$ is real and analytic, and it has the same radius of convergence as the matrix $\Theta$.

\end{document}